\documentclass[reprint,amsmath,amssymb,aps]{revtex4-2}
\usepackage{graphicx}
\usepackage{dcolumn}
\usepackage{bm}
\usepackage{float}
\usepackage{color}
\usepackage[dvipsnames]{xcolor}
\usepackage{hyperref}
\usepackage{graphicx}
\usepackage{braket}
\usepackage{appendix}
\usepackage{chemmacros}
\usepackage{natbib}
\usepackage{multirow}
\usepackage{amsmath}
\hypersetup{colorlinks=true, citecolor=blue, urlcolor=blue, linkcolor=blue}
\usepackage{soul}
\usepackage{mathrsfs}

\begin{document}

\preprint{APS/123-QED}

\makeatletter
\newcommand*{\balancecolsandclearpage}{%
  \close@column@grid
  \cleardoublepage
  \twocolumngrid
}
\makeatother
%\begin{document}

%\preprint{APS/123-QED} 

\

\title{
Deterministic role of chemical bonding in the formation of altermagnetism: Reflection from correlated electron system NiS
}   

\author{Arijit Mandal}\altaffiliation[arijitmandal1997@gmail.com]{}\affiliation{Condensed Matter Theory and Computational Lab, Department of Physics, IIT Madras, Chennai-600036, India}
\affiliation{Center for Atomistic Modelling and Materials Design, IIT Madras, Chennai-600036, India}
\author{Arindom Das}{}\affiliation{Condensed Matter Theory and Computational Lab, Department of Physics, IIT Madras, Chennai-600036, India}
\affiliation{Center for Atomistic Modelling and Materials Design, IIT Madras, Chennai-600036, India}    
%\email{satpathys@missouri.edu}
\author{B. R. K. Nanda}\altaffiliation[nandab@iitm.ac.in]{}\affiliation{Condensed Matter Theory and Computational Lab, Department of Physics, IIT Madras, Chennai-600036, India}
\affiliation{Center for Atomistic Modelling and Materials Design, IIT Madras, Chennai-600036, India}
%\email{nandab@iitm.ac.in}

%\date{\today}

\begin{abstract}
Altermagnetism, a new collinear magnetic state, has gained significant attention in the last few years, and the underlying mechanisms driving this quantum phase are still evolving. Going beyond the group theoretical analyses, which focus on providing a binary description of the presence or absence of the altermagnetic state, in this work, we explore the role of crystal chemical bonding. As the latter successfully integrates the crystal and orbital symmetries and is tunable, it provides a quantitative and realistic mechanism to explain the formation of altermagnetism. From the first principles calculations and tight-binding models within the framework of the linear combination of atomic orbitals on NiS,
we establish a set of selection rules for the formation of altermagnetism in the NiAs prototype compounds (e.g. CrSb, MnTe, etc.). Broadly, if single orbitals from Ni and S sites are involved in the bonding, the second neighbor interaction between the nonmagnetic atoms is a must to modulate the intra-sublattice interactions differently for the opposite spin sublattices so that the antiferromagnetic sublattice band degeneracy is lifted and momentum-dependent altermagnetic spin split (AMSS) appears. However, when multiple orbitals are involved from the Ni and S sites in the chemical bonding, altermagnetism is naturally present. Together, they amplify the AMSS.
Further, we propose twelve antinodal regions in the NiAs type hexagonal crystals, where AMSS split is maximum. Specific to NiS, AMSS increases with correlation, and for the edge valence and conduction bands, it can go beyond 1eV. The present study opens up new pathways to design chemical bonding driven selection rules in addition to the existing crystal symmetry criteria to tailor tunable altermagnetism.

\end{abstract}
\maketitle

\section{Introduction}

\begin{figure}[hbt!]
    \centering
   \includegraphics[width=1.0\linewidth]{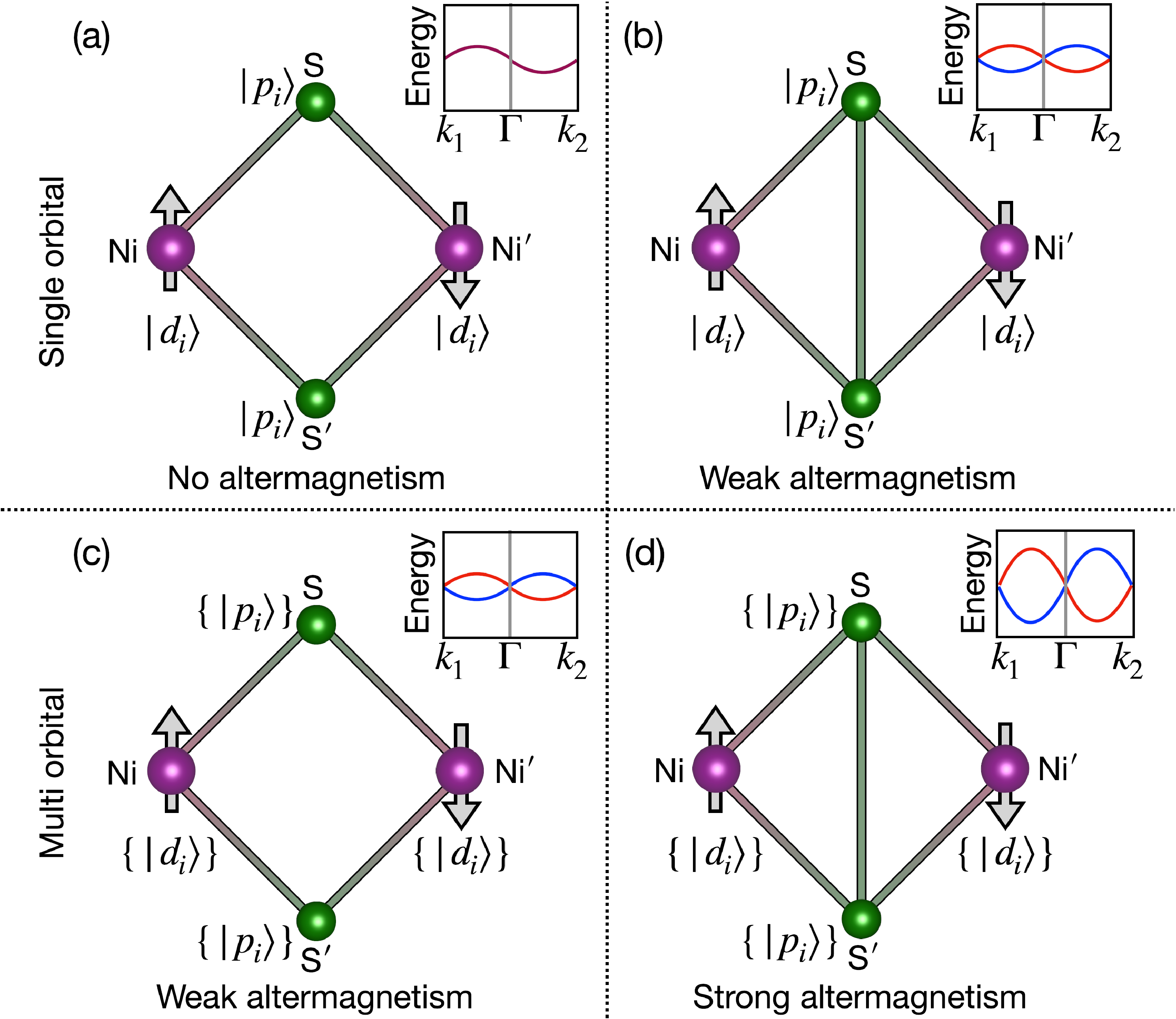}
    \caption{A schematic representation of the minimal interaction model that triggers the altermagnetism in NiS. The upper panel (a, b) represents the single orbital model, where one orbital each from the inequivalent atoms (Ni, Ni$^{\prime}$; S, S$^{\prime}$) form the basis. In the lower panel (c, d), multiple orbitals from each atom form the basis. The S-S$^{\prime}$ interactions are turned off on the left and turned on on the right. Ni and Ni$^{\prime}$ represent the two antiferromagnetic sublattices.}
    \label{schm}
\end{figure}

In the last few years, the subject of altermagnetism \cite{libor1, libor2} has become an emergent research topic in condensed matter physics. There is a gradual acceptance of this quantum state of matter as a new collinear magnetic phase, particularly distinguishable in the momentum space \cite{alter1, alter2, alter3, alter4, alter5, alter6, alter7, kerr1, ahe1}. It lies at the intersection of ferromagnetism and antiferromagnetism.  The altermagnet manifests a combined effect of ferromagnet and antiferromagnet. For example, the spins of nearest neighbor sites follow antiparallel alignment like antiferromagnet, but the time-reversal symmetry is not preserved universally in the momentum space \cite{alter1, alter2, alter3, alter4, alter5, alter6, alter7, kerr1, ahe1, libor1, libor2}. This nonrelativistic spin splitting with compensated magnetism is opening up a new window for gaining new fundamental insights into the world of complex many-body interactions involving lattice, charge, spin, and orbital degrees of freedom that the Bloch electrons possess in a crystalline system. Ideas also have been floated to expect a number of unconventional effects and transport, such as anomalous Hall effect \cite{ahe1, alter1, k-cl2}, spin Hall effect \cite{she1, she2, k-cl1}, nonlinear Hall effect \cite{nhe}, inverse spin Hall effect \cite{ishe}, magneto-optical Kerr effect \cite{kerr1, kerr2}, magnetoresistance \cite{magres1, magres2, magres3}, etc., to be exhibited by the altermagnets, and thereby, these compounds hold the promise for applications in futuristic quantum technologies.

A large fraction of research in altermagnetism is devoted to find the momentum-dependent spin-split of the bands in otherwise known antiferromagnets. These include rutile (RuO$_2$ \cite{alter1, alter2}, MnF$_2$ \cite{alter5}, MnO$_2$ \cite{mno2}), NiAs type (CrSb \cite{crsb, crsb-2, crsb-3}, MnTe \cite{mnte, mnte-2}), perovskite (SrRuO$_3$ \cite{kerr2}, CaCrO$_3$ \cite{cacro3}, LaMnO$_3$ \cite{lamno3}), FeSb$_2$ \cite{kerr1}, Mn$_5$Si$_3$ \cite{mn5si3}, $\kappa$-Cl \cite{k-cl1, k-cl2}, etc.. The compounds that stabilize in NiAs-type hexagonal crystal structure and exhibit antiferromagnetic spin ordering in real space are good examples of altermagnet with large band splitting ($\approx 1$ eV). Within this family of compounds the altermagnetism in CrSb and MnTe has widely been studied. The former shows metallic behavior, while the latter manifests insulating behavior. Another member of this family is the antiferromagnetic NiS, which was extensively studied in the past to understand correlation-driven physics. Specifically, its electronic state has been debated with some experiments predicting it as a metal \cite{metal-expt1, metal-expt2, metal-expt3, magnetic-moment} while others predicting it as an insulator\cite{insulator-expt1, insulator-expt2, insulator-expt3, insulator-expt4, insulator-expt5}. Theoretically, it is reported to be a good example for studying correlation driven metal to insulator transition (MIT) \cite{, metal-th1, metal-th2, metal-th3, insulator-th1, insulator-th2, insulator-th3, panda, self-consistent}.
%: Fig 1
%%%%%%%%%%%%%%%%%%%%%%%%%%%%%%%%%%%%%%%%%%%%%%%%%%%%%%%%
\begin{figure}[t]
    \centering
    \includegraphics[width=1\linewidth]{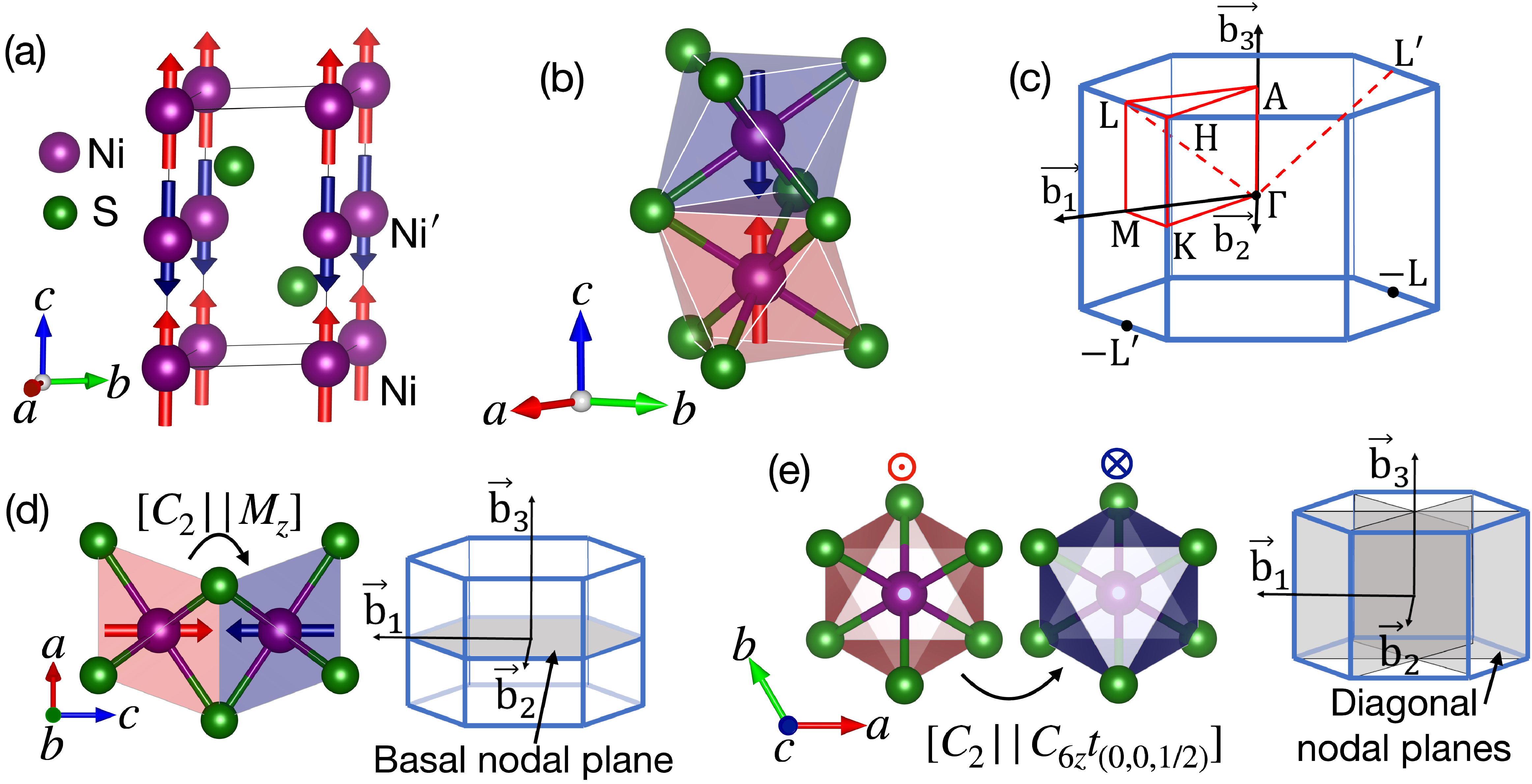}
    \caption{Crystal structure of NiS and symmetries associated with it. (a) The unit cell of the hexagonal crystal structure of NiS. The spins of the Ni atoms form an A-type antiferromagnetic order in the ground state. (b) Each Ni atom is surrounded by six S atoms to form a NiS$_6$ octahedron. The adjacent octahedra are face-shared. (c) The Brillouin zone of the hexagonal lattice. The high symmetric $k$-points are indicated. The conventional $k$-path demonstrating the antiferromagnetic degenerate bands is shown by the red solid line. The widely examined altermagnetic $k$-path is shown in a red dotted line.
     (d) and (e) represent the symmetry operations that connect two opposite spin sublattices. While $C_2$ acts on the spin space, $M_z$ and $C_{6z}t_{(0,0,1/2)}$ act on the real space. The $M_z$ gives one nodal plane and $C_{6z}t_{(0,0,1/2)}$ gives three nodal planes.
     Here, the symbols $\otimes$ and $\odot$ denote the spins into and out of the plane of the paper, respectively.
}
    \label{im1}
\end{figure}
%%%%%%%%%%%%%%%%%%%%%%%%%%%%%%%%%%%%%%%%%%%%%%%%%%%%%%%%

The primary objective of this work is to examine whether NiS exhibits altermagnetism like its isostructural counterparts, such as CrSb and MnTe and to identify the underlying mechanism driving the presence or absence of this non-trivial magnetic state. In most of the theoretical studies, symmetry analyses involving geometrical and magnetic space groups have been adopted to explain the cause of altermagnetism \cite{libor1, libor2}. However, we find that the insights from chemical bonding have hardly been explored in these studies. Both crystal symmetries and orbital symmetries together are determining factors for the direction-dependent chemical bonding which critically influences the band dispersion and localized spin-moment formation,  as well as exchange mechanisms driving the magnetic orderings. Therefore, insights from chemical bonding are crucial in developing a complete description of altermagnetism. At the time of writing this article, a perspective article by Fender et al. appeared in the literature, suggesting that the importance of chemical bonding it driving the altermagnetism should be investigated \cite{JACS}.
In this work, along with the density functional theory (DFT) calculations, we have employed a linear combination of atomic orbital approach (LCAO) as devised by Slater and Koster \cite{sktb}. In this approach, it is possible to intertwine the crystal and orbital symmetries.
Choosing NiS as a prototype also gives the freedom to study the effect of correlation and probably to realize the MIT in an altermagnet.

Our analysis throws up interesting insights, which are summarized in Fig. \ref{schm}. If the basis is formed by one orbital each from Ni and S atoms and the interactions are restricted to the nearest neighbor Ni-S paths only, the altermagnetism is not observed. The conventional antiferromagnetic sublattice (SL) band degeneracy (i.e., $\epsilon_{\uparrow}^{SL-1}(k)$ = $\epsilon_{\downarrow}^{SL-2}(k)$)is not lifted and the time-reversal symmetry is preserved. The degeneracy is lifted in the presence of the second neighbor interaction between the orbitals of the inequivalent nonmagnetic S atoms (Fig. \ref{schm}(b)). This second neighbor interaction modulates the nearest neighbor interactions for each of the antiferromagnetic sublattice, Ni - S/S$^{\prime}$ forming SL-1 and  Ni$^{\prime}$ - S/S$^{\prime}$ forming SL-2, differently. As a result, there is a split between $\epsilon_{\uparrow}^{SL-1}(k)$ and $\epsilon_{\downarrow}^{SL-2}(k)$ which gives rise to altermagnetism. The second neighbor interactions among the equivalent atoms (Ni-Ni, Ni$^{\prime}$-Ni$^{\prime}$, S-S, and S$^{\prime}$-S$^{\prime}$) do not break the band degeneracy. The differently modulated nearest neighbor interactions can also be realized in the absence of S-S$^{\prime}$ interaction, provided multiple orbitals from both Ni and S sites form the basis (Fig. \ref{schm}(c)). While choosing the orbitals from S-atoms in the multi orbital basis framework, we find that the  S-p$_z$ orbital has a significant role for driving the altermagnetism. Collectively, both of these factors create strong altermagnetism (Fig. \ref{schm}(d)) in Ni-S and its isostructures, such as CrSb and MnTe.

\section{Crystal Structure and Computational Details}
The compound NiS stabilizes in a hexagonal crystal structure with space group $P6_3/mmc$ as shown in Fig. \ref{im1}(a) with Ni and S occupying the Wyckoff positions $2a$ and $2c$, respectively. The experimental lattice parameters for this hexagonal system are $a=b=3.437 \AA$ and $c=5.345 \AA$ \cite{crystal1, crystal2}, and the same have been used for the DFT calculations. Since we intend to examine the momentum-dependent splitting of the sublattice bands, which, as per the literature, is strongly crystal symmetry dependent, we find it important to highlight the following symmetry operations that this compound possesses. This compound belongs to dihexagonal dipyramidal crystal class ($\frac{6}{m}$ $\frac{2}{m}$ $\frac{2}{m}$; $D_{6h}$). It has an inversion centre and a screw axis with a six-fold rotation followed by a half translation along $c$. Moreover, it has a glide plane parallel to the $c$ axis. The Wyckoff position $2a$ and $2c$ have the site symmetry $\Bar{3}m$ and $\Bar{6}2m$. 

The salient features of NiS is that
it has NiS$_6$ octahedra, which are face-shared. Furthermore, the nearest neighbor Ni-S-Ni bond angles are $\approx 90^{\circ}$ on the $ab$ plane and $\approx 68^{\circ}$ along $c$ axis. As mentioned earlier, NiS stabilizes in a $A$-type antiferromagnetic order with a N\'eel temperature of  $265$K. The consensus in the literature is that the $90^{\circ}$ bond angle drives a weak ferromagnetic ordering via superexchange interaction in the $ab$ plane, while the $68^{\circ}$ bond angle along the $c$ axis makes the superexchange interaction antiferromagnetic \cite{panda}. 

In this study, the DFT calculations are carried out using the plane-wave-based projector augmented wave (PAW) method \cite{paw1, paw2}, as implemented in the Vienna \textit{ab initio} Simulation Package (VASP) \cite{vasp} to examine the electronic and magnetic structure of the compound. The Perdew-Burke-Ernzerhof generalized gradient approximation (PBE-GGA) is employed for the exchange-correlation functional. A $\Gamma$ centred $12\times12\times8$ k-mesh has been used to integrate the Brillouin zone (BZ). The energy cut-off for the plane wave basis set is taken to be $500$ eV. To account for the strong correlation effects, the DFT + U method has been used with the effective Hubbard parameter $U_{\text{eff}} = U - J$ in the rotationally invariance framework as provided by Dudarev \cite{dudarev}. Following the DFT calculations, a minimal basis tight-binding (TB) model Hamiltonian is developed to build a relation between momentum vector $k$ and spin split band.

\section{Description of the ground state electronic structure}
\begin{figure}[t]
    \centering
   \includegraphics[width=1\linewidth]{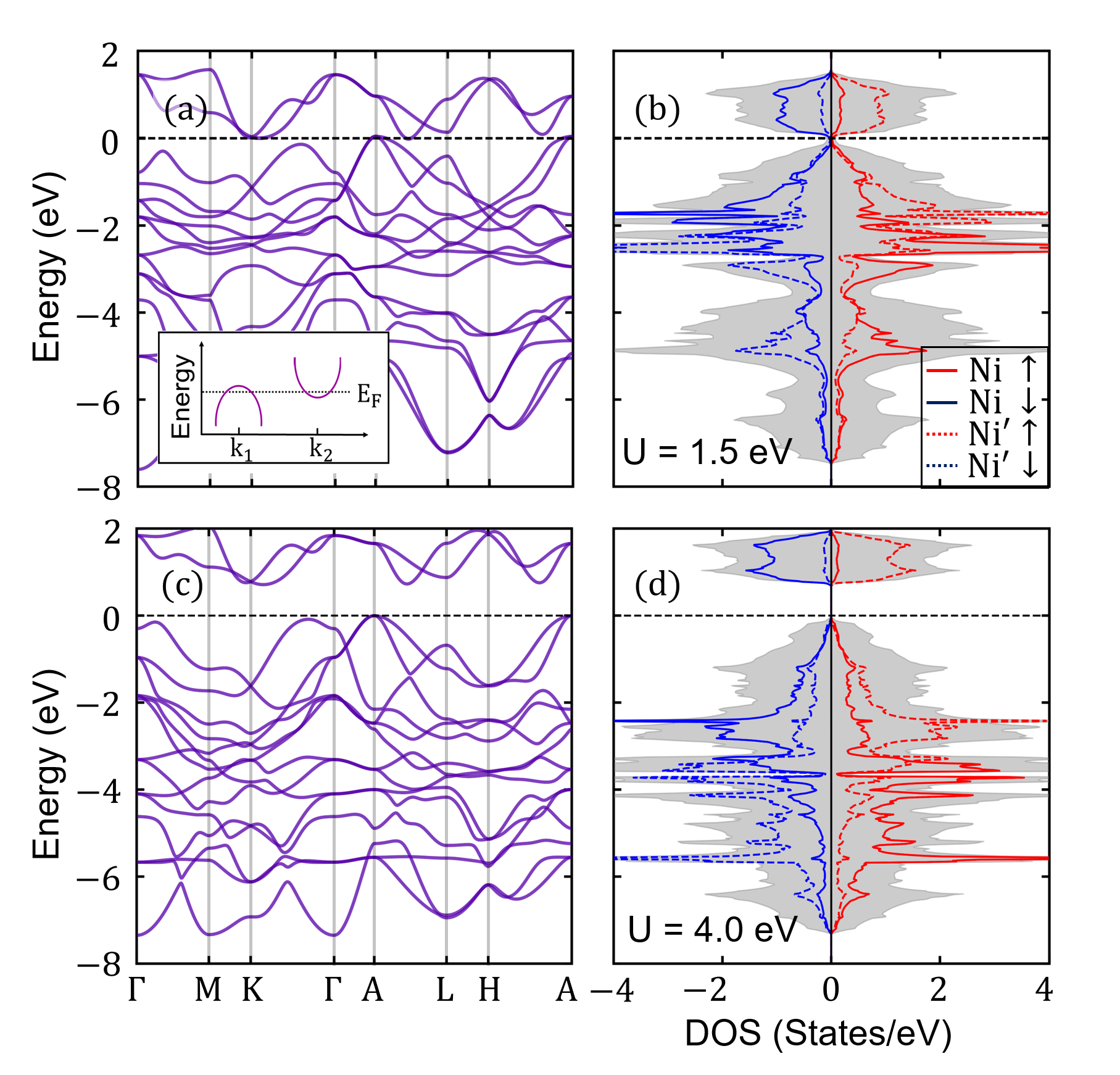}
    \caption{Band structure (left) and DOS (right) of antiferromagnetic NiS for $U=1.5$ eV ((a) and (b)) and $U=4$ eV ((c) and (d)). The area filled with gray color in (b) and (d) represents the total DOS of spin-up and spin-down states. In the inset of (a), we have schematically shown the presence of electron and hole pocket. Later, we will see that the $k$-points along this conventional $k$-path lie on the nodal plane or on the BZ boundary. Therefore, antiferromagnetic sublattice band degeneracy is maintained.}
    \label{im2}
\end{figure}

Here we briefly discuss the electronic structure and stabilization of antiferromagnetic ground state. In the nonmagnetic configuration and in the absence of proper correlation strength, the Ni-$d$ states ($t_{2g}$ components) dominate the vicinity of Fermi level in the valence band spectrum, while the S-$p$ bands lie below the Ni-$d$ bands. In the conduction band spectrum, both Ni-$d$ ($e_g$ components) and S-$p$ states form the bottom conduction band (see Fig. \ref{NM_U=0} in Appendix \ref{nonmagnetic}). With the inclusion of $U$ and stabilization of the antiferromagnetic ordering, the valence Ni-$d$ states are pushed down in the energy level, and S-$p$ states dominate the vicinity of the Fermi level. The antiferromagnetic band structure for $U=1.5$ and $4$ eV, which lie on either side of critical $U$ at which the theoretically proposed MIT occurs, are shown in Fig. \ref{im2} \cite{panda}. Along the conventional $k$-path ($\Gamma-M-K-\Gamma-A-L-H-A$), the spin majority/minority bands of the opposite sublattices coincide as already been reported. It is interesting to note that, in the case of $U=1.5$ eV, at two specific $k$-points ($K$ and a point between $A$ and $L$), we observe the bottom conduction band forming a tiny electron pocket, while at the high symmetry point $A$, a hole pocket is formed. In the rest of the BZ, there is a well-defined separation between the conduction and valence bands. These electron and hole pockets vanish with increasing values of $U$ and such a band transition is described as an MIT transition. A detailed analysis shows that even in the nonmagnetic configuration ($U=0$), there is a depleted density of states (DOS) at the Fermi energy ($E_F$) (see Appendix \ref{nonmagnetic}, Fig. \ref{NM_U=0}) as at very few $k$-points the bands crosses the $E_F$. Around $U=1$ to $1.5$ eV, the localization of bands and formation of local spin moments occur. This leads to the formation of the pockets. Therefore at the DOS of Fig. \ref{im2}(b) shows the system achieves a semimetallic instead of a normal metallic phase. As expected, the higher $U$ stabilizes a correlated insulating phase.

%:Fig 2

\section{Altermagnetism in $\textbf{NiS}$}
\begin{figure}[t]
    \centering
   \includegraphics[width=1\linewidth]{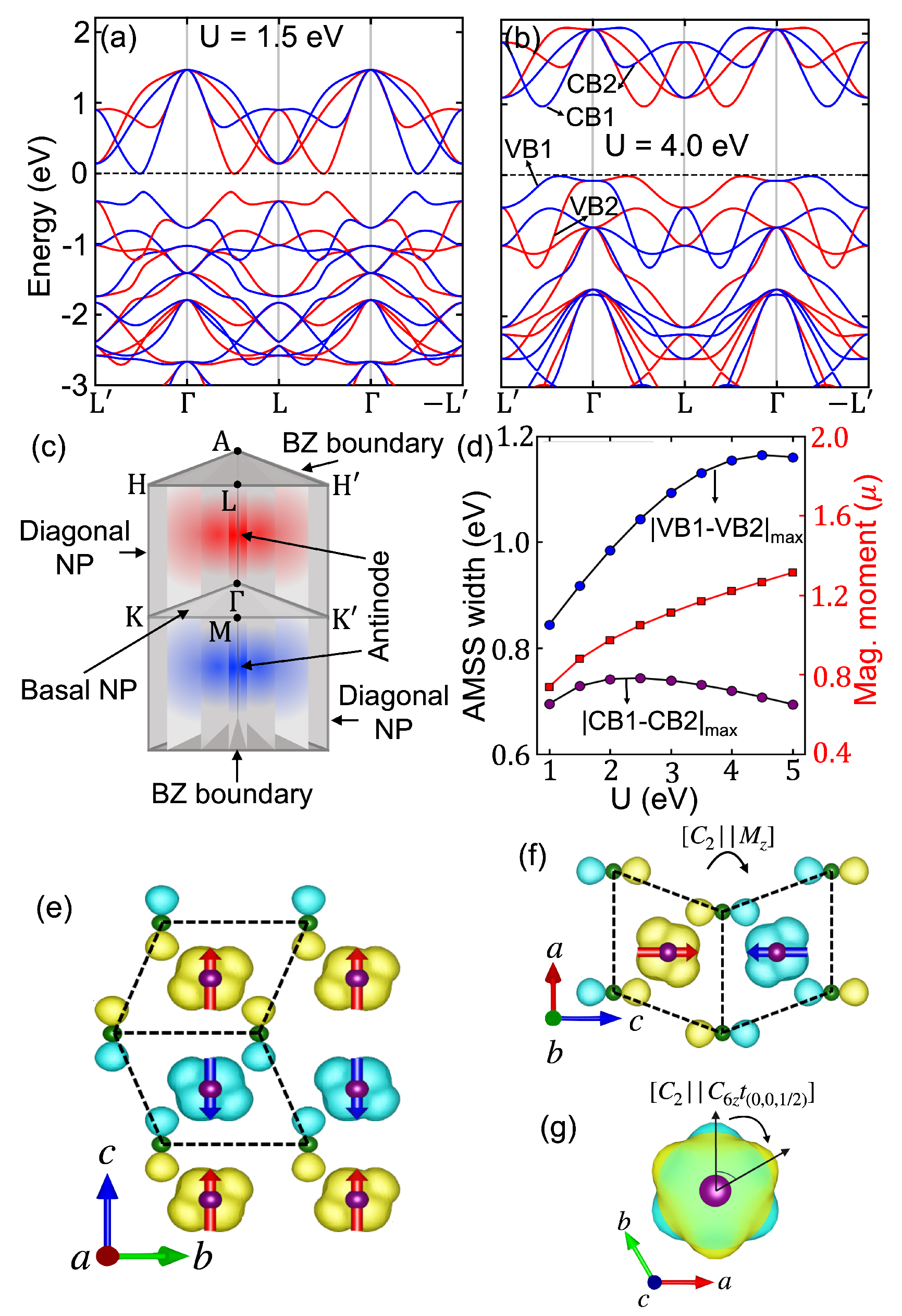}
    \caption{The nonrelativistic momentum dependent spin splitting along one of the altermagnetic $k$-path with (a) $U = 1.5$ eV and (b) $U = 4$ eV. (c) A schematic representation of antinodes between two adjacent nodal planes. As we move from one nodal plane to the other, the AMSS starts increasing, becomes maximum around the antinodes, and then starts decreasing to become zero at the nodal plane. As we cross a nodal plane, the bands of the opposite sublattices flip their position, which is a characteristic feature of altermagnetism. (d) The strength of maximum AMSS for a pair of altermagnetic bands as a function of onsite correlation parameter $U$. To demonstrate the dependence, we have taken a valence band and a conduction band pair, as indicated by VB1, VB2 and CB1, CB2, respectively, in (b). (e)-(g) represent the spin densities projected on the $bc$ (isosurface value 0.005), $ac$ (isosurface value 0.005), and $ab$ (isosurface value 0.1) planes.}
    \label{im3}
\end{figure}

As discussed earlier, NiS has a lot of similarities with CrSb and MnTe. They are isostructures and they stabilize in $A$-type antiferromagnetic ordering. Like NiS, the electronic structure of MnTe is strongly correlated and its ground state is an insulator. Theoretical studies have predicted strong altermagnetic spin split (AMSS) in CrSb and MnTe. By plotting the bands along the altermagnetic $k$-path, as shown in Fig. \ref{im3}(a) and (b), we observe a comparable or even larger AMSS. Therefore, NiS is an ideal prototype to gain insight into the underlying mechanism driving the altermagnetism in the family of transition metal based hexagonal binary compounds with space group $P6_3/mmc$. It also helps us in understanding the effect of strong correlation on altermagnetism.

%:Fig 3
%%%%%%%%%%%%%%%%%%%%%%%%%%%%%%%%%%%%%%%%%%%%%%%%%%%%%%%%%%%%%

The first step is to analyze the band structure along the altermagnetic $k$-path $L^{\prime}-\Gamma-L-\Gamma-(-L^{\prime})$. From Fig \ref{im3}, we observe that the altermagnet band splitting is maximum for the bands around the $E_F$ and gradually reduces as we move away deep inside the valence and conduction band sections. In the semimetallic phase, the strength of AMSS is around $1$ eV. As the system achieves an insulating state in the strong correlation, the splitting further increases.

In addition to the above basic observations, we further identified the following important salient features. There are four nodal planes, as shown in Fig. \ref{im1}(e) and (f). At any point on these nodal planes as well as the BZ boundaries, the bands behave like perfect antiferromagnetic without any splitting of the sublattice bands. On $k$-paths lying in any other plane passing through $\Gamma$, the AMSS is observed. The AMSS increases as we move away from a given nodal plane and it reaches a maximum at the middle and further decreases as we move to the adjacent nodal plane. This sinusoidal pattern of AMSS gives rise to $12$ antinodal points around which maximum AMSS occurs. Two of them lying in the middle of two adjacent diagonal nodal planes and separated by the basal nodal plane are shown in figure \ref{im3}(c). The color red and blue indicate the spin flip in the AMSS. 

Interestingly, we find that the AMSS width varies with $U$ and there exists a competition between magnetization and localization in determining the AMSS strength (see Fig. \ref{im3}(d)). Let us first focus on the AMSS for the bottom two conduction bands marked as CB1 and CB2. We find that it first increases with $U$ up to $U \approx 2.5$ eV and then monotonically decreases. The magnetization rapidly increases for the initial range of $U$ and this enhanced magnetization strengthen the AMSS. This is in agreement with the earlier observation by Chakraborty et al. \cite{sinova-U}. However, as the narrowing down the bands occurs for higher value of $U$ (see Figs. \ref{im2}(a), (c), and \ref{im3}(a), (b)), the scope for larger AMSS driven by the magnetization is constrained and therefore the AMSS starts decreasing. In the case of top two valence bands marked as VB1 and VB2, it is little more complex as the localization of the bands as a function of $U$ is poor. The bandwidth remains nearly constant or it increases, albeit small, with $U$. This is due to the fact that the Ni-$d$ dominated valence bands penetrate the S-$p$ dominated valence band spectrum as the former lowers their energy with U. The orbital intermixing complicates the many-body interaction further. In an earlier work in a different context \cite{chain-dop}, we have shown that the bandwidth can increase with U in a multiorbital picture. In this scenario, the enhanced magnetization further enhances the AMSS. We observe that, the AMSS saturates to $\approx 1.2$ eV as the value of $U$ goes beyond $4$ eV. Such a large splitting makes this system ideal for experimental validation and future applications. 

In a couple of seminal works, \v{S}mejkal et al. \cite{libor1, libor2} used group theoretical analysis to explain the origin of altermagnetism. According to them, in altermagnet crystals, there must exist a coset decomposition of the parent point group $\boldsymbol{G}$ through a halving subgroup $\boldsymbol{H}$ that contains the symmetry operations within the same spin sublattices. The symmetry element of the coset $\boldsymbol{G}-\boldsymbol{H}=A\boldsymbol{H}$ makes real space transformation such that the atoms between the opposite spin sublattices are interchanged. Here, $A$ has to be a proper or improper rotation but not an inversion. The group $\boldsymbol{G}$ and $\boldsymbol{H}$ form the Laue spin group $\boldsymbol{R}_S = [E||\boldsymbol{H}] + [C_2||\boldsymbol{G}-\boldsymbol{H}]$. In the non-relativistic framework, the magnet with this spin group has a spin split band structure, which is the characteristic feature of an altermagnet. For NiS and its isostructure, $\boldsymbol{G}$ is $6/mmm$, $\boldsymbol{H}$ is $\bar{3}m$. The six-fold screw symmetry operation $C_{6z}$ defines $A$. The spin density plotted in Fig. \ref{im3}(e)-(g) reflects the following two salient features. (i) The spin density of the two sublattices, formed by Ni and Ni$^{\prime}$, are connected through mirror reflection (see Fig. \ref{im3}(f)) as well as through six-fold rotation (see Fig. \ref{im3}(g)). This is along the line that in every altermagnetic material, the anisotropic magnetisation densities of the two sublattices are connected by the existing mirror and rotational symmetries of the corresponding crystal. Different spin density of individual sublattices leads to differences in the nature of the chemical bonding of each sublattice, which will be discussed later. As the chemical bonding differs, the sublattice resolved band dispersion in the momentum space relation differs, which gives rise to AMSS. (ii) It is interesting to note that both Ni and Ni$^{\prime}$ induce the magnetization in S and S$^{\prime}$. Further, the acquired spin density of S/S$^{\prime}$ on the Ni side is equal and opposite to that on the Ni$^{\prime}$ side. As a result, the S atoms as a whole are nonmagnetic.

\section{Origin of Altermagnetism: A Model Hamiltonian Analysis}
\label{TB-sec}
Analysis from the group theory provides a binary description of either the possible presence or absence of altermagnet in the collinear magnetic systems. A quantitative description, such as the strength of the AMSS, requires a more nuanced approach which should be capable of providing tunable parameters and a wider set of deterministic criteria. Since chemical bonding has the most significant role to explain the electronic and magnetic structure of a system, it must be taken into account while developing models and hypotheses. In this section, we present a TB model using the formalism of LCAO formalism as devised by Slater and Koster. 

Several TB Hamiltonians have been proposed where an additional coupling has been introduced to initiate AMSS. For example, Roig et al. included generic inter and intra-sublattice hopping in addition to the sublattice independent hopping to provide a minimal model stabilizing altermagnetism in systems with different space group and point group symmetries \cite{andreas-alter}. On the other hand, Sato et al. introduced a phase-modulated second neighbor electron hopping to describe altermagnetism in general in hexagonal systems \cite{brink}. Several other model Hamiltonians are also proposed \cite{ahe1, hayami-model, lieb-lattice-model, weak-coupling-model}. However, as we understand none of them have considered the real orbitals in the basis. In contrast to these formalisms, the LCAO model presents a more realistic description as it brings orbital-driven chemical bonding through Slater-Koster (SK) matrix elements. Therefore, it intertwines the crystal and orbital symmetries.

The standard SKTB Hamiltonian in the second quantization formalism in a standard form can be written as:
\begin{eqnarray}
    \mathscr{H} &=& \sum_{i, \mu, \tau, \sigma} \epsilon_{i\mu\tau\sigma} c_{i\mu\tau\sigma}^\dagger c_{i\mu\tau\sigma} \nonumber \\
    &\mp& \sum_{i, \mu, \tau, \sigma}(-1)^{\tau} \Delta_{i, \mu, \tau, \sigma}/2 c_{i\mu\tau\sigma}^\dagger c_{i\mu\tau\sigma}
    \nonumber \\
    &+& \sum_{i, j, \mu, \nu, \sigma} (t_{i\mu j\nu \sigma} c_{i\mu \sigma}^\dagger c_{j\nu \sigma} + \text{H.c.}),
    \label{moel_ham}
\end{eqnarray}
 where, $i,\ j$ are site indices, $\mu,\ \nu$ depict orbital indices, and $\sigma$ describes the spin indices. $\tau = 1$ represents the sublattice where spin-up states form the majority spin channel, and $\tau = 2$ represents the sublattice where spin-down states form the majority spin channel. We consider $d$-orbitals of Ni and $p$-orbitals of S for constructing the basis of the Hamiltonian.  In the above equation, the first term represents the onsite energies of the individual orbitals. The second term, $\Delta/2$, represents Hund's coupling of the spin-polarized orbitals (Ni $d$-orbitals in the present case). The orbitals of two sublattices, Ni ($\tau = 1$) and Ni$^{\prime}$ ($\tau = 2$), have equal and opposite values of $\Delta/2$ as the system is in $A$-type antiferromagnetic configuration. The third term represents the electron hopping integrals, which, as discussed earlier, are expressed using the SK relations. To demonstrate how the matrix elements are formed out of the hopping integrals, in appendix \ref{hopping}, we have explicitly derived the expressions for the elements $\langle Ni-d_{xy}|H|S-p_{x}\rangle$, $\langle Ni'-d_{xy}|H|S-p_{x}\rangle$, and $\langle S-p_{x}|H|S'-p_{x}\rangle$. The rest of the matrix elements can be constructed in a similar way.

 % %%%%%%%%%%%%%%%%%%%%%%%%%%%%%%%%%%%%%%%%%%%%%%%%%%%%%%%%%%%
%Fig 6
\begin{figure}[hbt!]
    \centering
   \includegraphics[width=1\linewidth]{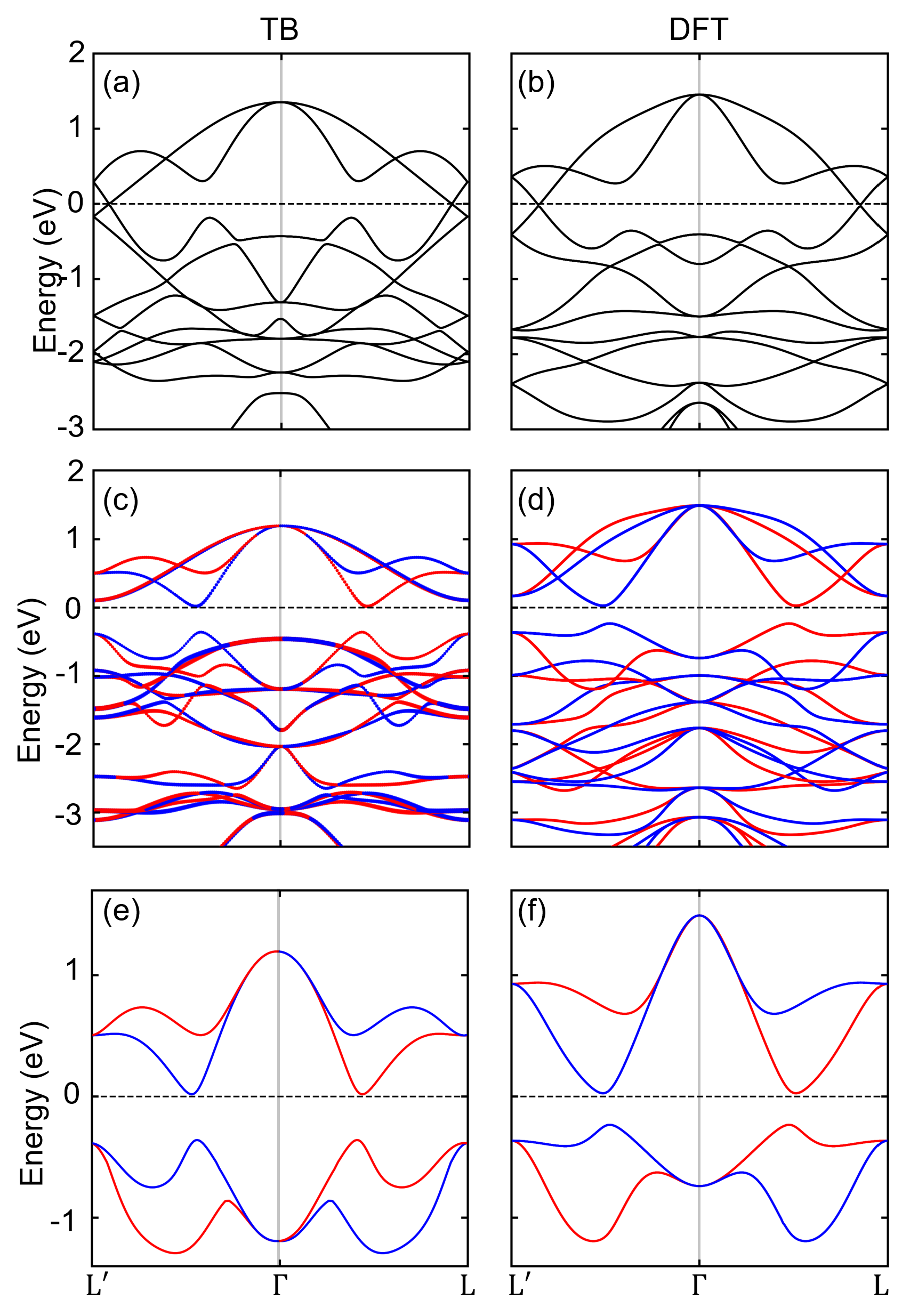}
    \caption{The nonmagnetic and altermagnetic band structure of NiS as obtained from the TB model (left panel) and DFT (right panel). (a) and (b) represent the nonmagnetic band structure, while (c) and (d) represent the altermagnetic band structure. For DFT, the value of $U$ is taken to be $1.5$ eV and for the TB model, the value of $\Delta_{\text{Ni}}$ is taken to be $1.5$ eV. In the last row ((e) and (f)), a couple of altermagnetic bands are highlighted for clarity.}
    \label{TB_NM}
\end{figure}

% 
%%%%%%%%%%%%%%%%%%%%%%%%%%%%%%%%%%%%%%%%%%%%%%%%%%%%%%%%%%%
\begin{table*}[hbt!]
\begin{center}
\caption{Estimated values of various hopping interaction strength and onsite energies in eV unit. The $\epsilon_{\text{Ni}}$ and $\epsilon_{\text{S}}$ are the onsite energies of Ni-$d$ and S-$p$ orbitals, respectively. $t_{n(s)\sigma 1}$ and $t_{n(s)\pi 1}$ represent the nearest-neighbor Ni(S)-Ni(S) $\sigma$ and $\pi$ interactions, respectively, while $t_{n(s)\sigma 2}$ and $t_{n(s)\pi 2}$ represent the same for nearest-neighbor Ni(S)-Ni$^{\prime}$(S$^{\prime}$) interactions. $t_{n\sigma 3}$ and $t_{n\pi 3}$ are the $\sigma$ and $\pi$ interactions between second nearest neighbor Ni-Ni$^{\prime}$ atom. $t_{ns\sigma}$ and $t_{ns\pi}$ denote $\sigma$ and $\pi$ interaction between different types of Ni and S atoms.}
\setlength{\tabcolsep}{6pt}
\renewcommand{\arraystretch}{1.5}
\begin{tabular}{c c| c c| c c c c| c c| c c| c c}
\hline
\hline
\multicolumn{2}{c|}{Onsite energy} & \multicolumn{2}{c|}{Ni-Ni interaction} & \multicolumn{4}{c|}{Ni-Ni$^{\prime}$ interaction} & \multicolumn{2}{c|}{S-S interaction} & \multicolumn{2}{c|}{S-S$^{\prime}$ interaction} & \multicolumn{2}{c}{Ni-S interaction}\\
$\epsilon_{\text{Ni}}$ & $\epsilon_{\text{S}}$ & $t_{n\sigma 1}$ & $t_{n\pi 1}$ & $t_{n\sigma 2}$ & $t_{n\pi 2}$ & $t_{n\sigma 3}$ & $t_{n\pi 3}$ &  $t_{s\sigma 1}$ & $t_{s\pi 1}$ &  $t_{s\sigma 2}$ & $t_{s\pi 2}$ & $t_{ns\sigma}$ & $t_{ns\pi}$\\
\hline
$-1.85$ & $-4.69$ & $0.116$ & $-0.027$ & $-0.549$ & $0.436$ & $-0.3$ & $0.032$ & $0.519$ & $-0.075$ & $1.112$ & $-0.325$ & $-1.42$ & $0.42$\\
\hline
\hline
\end{tabular}
\label{int-tab1}
\end{center}
\end{table*}
%%%%%%%%%%%%%%%%%%%%%%%%%%%%%%%

 As an initial step, we first validated the TB model for the nonmagnetic configuration, where only the first and third terms are required. We have restricted the interactions up to first neighbor in the plane ($d=3.43 \AA$) and up to second neighbor along the out-of-plane ($d_{max}=4.4 \AA$). This gives us $12$ hopping parameters ($t$s) as listed in table \ref{int-tab1}. In addition, there are two onsite parameters ($\epsilon_{Ni}$ and $\epsilon_S$). The optimized values of these parameters (see table \ref{int-tab1}) and the resulting TB bands are obtained, which is shown in Fig. \ref{TB_NM}(a), which agrees reasonably well with the DFT obtained band structure (see Fig. \ref{TB_NM}(b)). After the validation, the antiferromagnetic Hamiltonian is designed using the hopping interaction strengths of the nonmagnetic case, where the hopping interaction between two majority (and minority) spin channels of the opposite Ni sublattices is prohibited. %The interaction between the majority and minority spin channels is approximately screened by a factor of $\Delta$. 
 The resultant band structure is shown in Fig. \ref{TB_NM}(c). It successfully reproduces the DFT-obtained AMSS feature along the k-path $L'-\Gamma-L$. The band structure looks slightly different as compared to the DFT+U obtained altermagnetic band structure of Fig. \ref{TB_NM}(d) since our model Hamiltonian does not consider the Hubbard interaction. However, within the DFT + mean field approximation, U and $\Delta$ are of the same order. To identify the deterministic role of chemical bonding in driving the AMSS, below, we will discuss a series of specific cases.\\

\textbf{Case I- Single orbital from each of the nickel and sulphur atoms; S-S$^{\boldsymbol{\prime}}$ interaction is absent (Fig. \ref{schm}(a)):}
As discussed earlier, there are two equivalent Ni atoms (Ni and Ni$^{\prime}$) and two equivalent S atoms (S and S$^{\prime}$). This makes a $8\times 8$ Hamiltonian matrix by taking both the spins into account. Since electrons cannot hop between two opposite spin channels in the absence of spin-orbit coupling, the matrix is block diagonal with two $4 \times 4$ blocks $\mathcal{H}_{4 \times 4}^{\uparrow \uparrow}$ and $\mathcal{H}_{4 \times 4}^{\downarrow \downarrow}$.

In a minimal interaction model, the Ni-Ni$^{\prime}$ interaction can be ignored due to the following points. (a) This is a second neighbor interaction (Ni-S being the nearest) and hence weak. (b) The strength of the interaction is further screened approximately by a factor $t/\Delta$. The latter is due to the fact that the energy separations of $d$-orbitals in the same spin channel between the two sublattices are $\Delta$. Further, if we block the S-S$^{\prime}$ interactions as well as the interactions between the equivalent atoms (Ni(Ni$^{\prime}$)-Ni(Ni$^{\prime}$) and S(S$^{\prime}$)-S(S$^{\prime}$)), the resultant block matrices with the basis set in the order $|\text{Ni}-d \rangle$, $|\text{Ni}^{\prime}-d \rangle$, $|\text{S}-p \rangle$, $|\text{S}^{\prime}-p \rangle$, takes the general shape:
\begin{eqnarray}
%\begin{align}
    &&\mathcal{H}_{4 \times 4}^{0\uparrow \uparrow(\downarrow \downarrow)} = \nonumber \\
    && \begin{pmatrix}
        -(+)\frac{\Delta_{\text{Ni}}}{2} + \alpha & 0 & w_1 + iw_2 & w_3 + iw_4\\
        0 & +(-)\frac{\Delta_{\text{Ni}}}{2} + \alpha & v_1 + iv_2 & v_3 + iv_4 \\
        w_1 - iw_2 & v_1 - iv_2 & 0 & 0 \\
        w_3 - iw_4 & v_3 - iv_4 & 0 & 0
    \end{pmatrix}.
\label{upblock}
%\end{align}
\end{eqnarray}
Here, $\alpha$ is the difference between free atomic onsite energies of Ni-$d$ and S-$p$ orbitals.
To make it simple, we have taken the free atomic onsite energies of the orbitals to zero.
The characteristic equation for both spin-up and spin-down blocks is a quartic function and  can respectively be written as:
\begin{eqnarray}
    \lambda_{\uparrow}^4 + b_{\uparrow}(k)\lambda_{\uparrow}^3 + c_{\uparrow}(k)\lambda_{\uparrow}^2 + d_{\uparrow}(k)\lambda_{\uparrow} + e_{\uparrow}(k) &=& 0 \nonumber \\
    \lambda_{\downarrow}^4 + b_{\downarrow}(k)\lambda_{\downarrow}^3 + c_{\downarrow}(k)\lambda_{\downarrow}^2 + d_{\downarrow}(k)\lambda_{\downarrow} + e_{\downarrow}(k) &=& 0 \nonumber \\
    \label{eigeneqn}
\end{eqnarray}

In table \ref{int-tab6}, we have listed the relation between the matrix elements of Eq. \ref{upblock} for different Ni-S pair orbital interactions. There primarily exist two sets of relations, type I and type II, which are expected to influence the coefficients of characteristic Eq. \ref{eigeneqn}. 

\begin{table}
\begin{center}
\caption{Relation between the matrix elements, $w$s and $v$s of the Hamiltonian of Eq. \ref{upblock} for different Ni-S pair orbital interactions. Irrespective of the orbitals forming the basis, $w_1^2 + w_2^2= v_1^2 + v_2^2$}
\setlength{\tabcolsep}{6pt}
\renewcommand{\arraystretch}{2}
\begin{tabular}{p{4cm} | p{4cm}}
\hline
\hline
 \textbf{Type I} & \textbf{Type II} \\
 $(w_3 + i w_4) = (w_1 + i w_2)^{\ast}$ & $(w_3 + i w_4) = -(w_1 + i w_2)^{\ast}$ \\
 $(v_3 + i v_4) = (v_1 + i v_2)^{\ast}$ & $(v_3 + i v_4) = -(v_1 + i v_2)^{\ast}$ \\
\hline
$d_{xy}$-$p_y$, $d_{xz}$-$p_z$, $d_{x^2-y^2}$-$p_x$, $d_{z^2}$-$p_x$, $d_{xy}$-$p_z$, $d_{yz}$-$p_x$, $d_{xz}$-$p_y$ & $d_{xy}$-$p_x$, $d_{yz}$-$p_z$, $d_{x^-y^2}$-$p_y$, $d_{yz}$-$p_y$, $d_{xz}$-$p_x$, $d_{x^2-y^2}$-$p_z$, $d_{z^2}$-$p_y$, $d_{z^2}$-$p_z$\\
\hline
\hline
\end{tabular}
\label{int-tab6}
\end{center}
\end{table}

\begin{figure}[hbt!]
    \centering
   \includegraphics[width=1.0\linewidth]{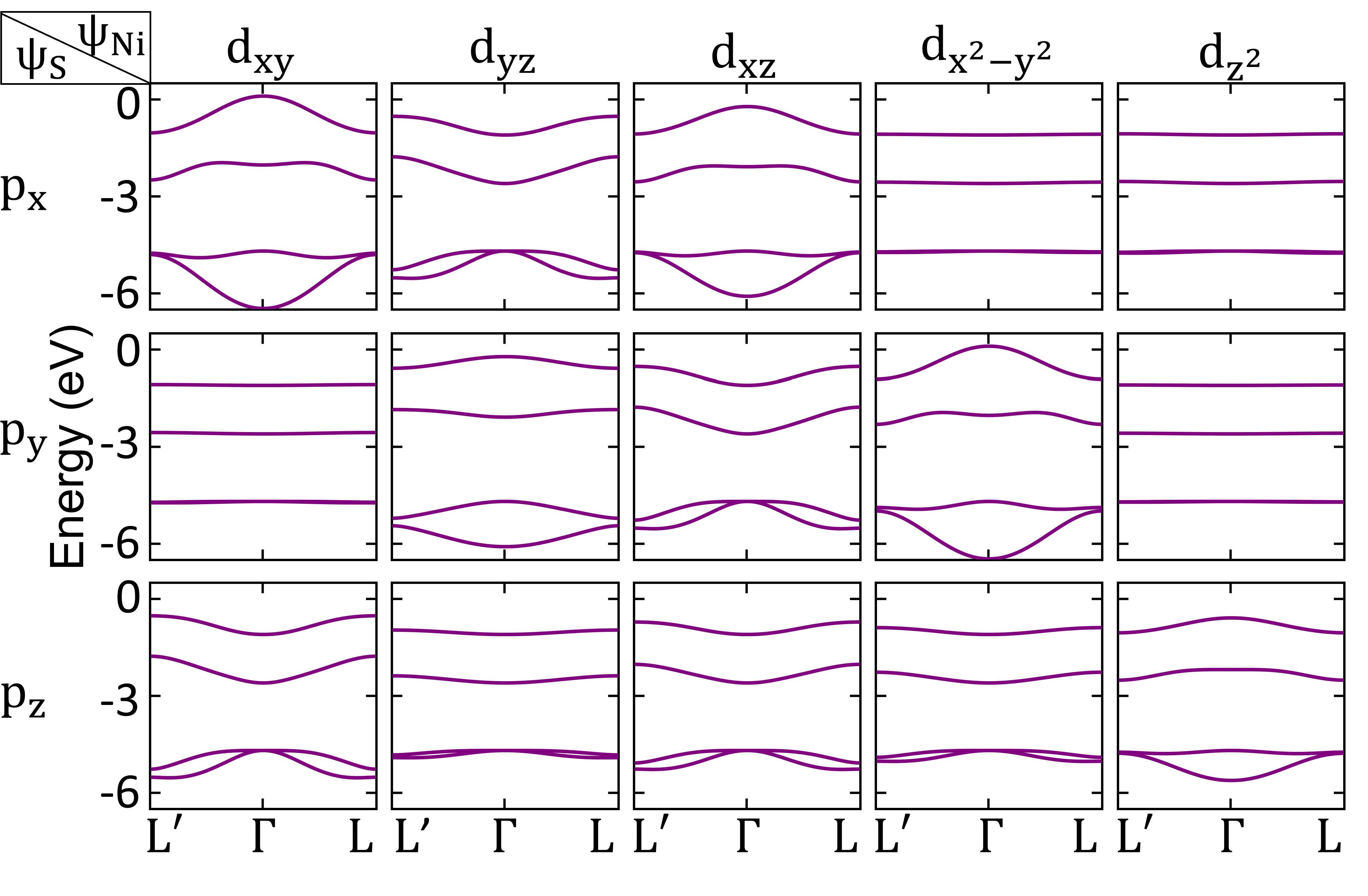}
    \caption{The band structure obtained from the single orbital TB model corresponding to Eq. \ref{upblock}. It infers that no altermagnetism will be observed in the single orbital model in the absence of S-S$^{\prime}$ interaction. Each of the figures corresponds to a pair $d$-$p$ orbital as indicated.}
    \label{im10}
\end{figure}

In the present case, due to the antiferromagnetism nature of the Hamiltonian, the coefficients of $\lambda^3$ ($b_{\uparrow}$ and $b_{\downarrow}$) are found to be zero for each of the block. The coefficient of $\lambda^1$ are derived to be $d_{\uparrow(\downarrow)} = +(-) 8 \Delta_{\text{Ni}} ( (w_1^2 + w_2^2)-(v_1^2 + v_2^2))$. Furthermore, the SK relations ensures that $w_1^2 + w_2^2 = v_1^2 + v_2^2$, $d_{\uparrow(\downarrow)} = 0$ (see  Appendix \ref{hopping}). Therefore, the variation in the spin-up and spin-down eigenvalues is now solely dependent on the coefficients of $\lambda^2$ ($c_{\uparrow(\downarrow)}$) and $\lambda^0$ ($e_{\uparrow(\downarrow)}$). As expressed in the Eq. \ref{coeff-eq2}, \ref{coeff-eq4}, \ref{coeff-eq6}, and \ref{coeff-eq8} of appendix \ref{hopping}, they are found to be identical for the whole BZ for both spin-up and spin-down blocks (consider $u_1$ and $u_2$ to be zero). As a result, the Hamiltonian of Eq. \ref{upblock} produces a perfect antiferromagnetic band structure which is also reflected in Fig. \ref{im10}.\\

\begin{figure}[hbt!]
    \centering
   \includegraphics[width=1\linewidth]{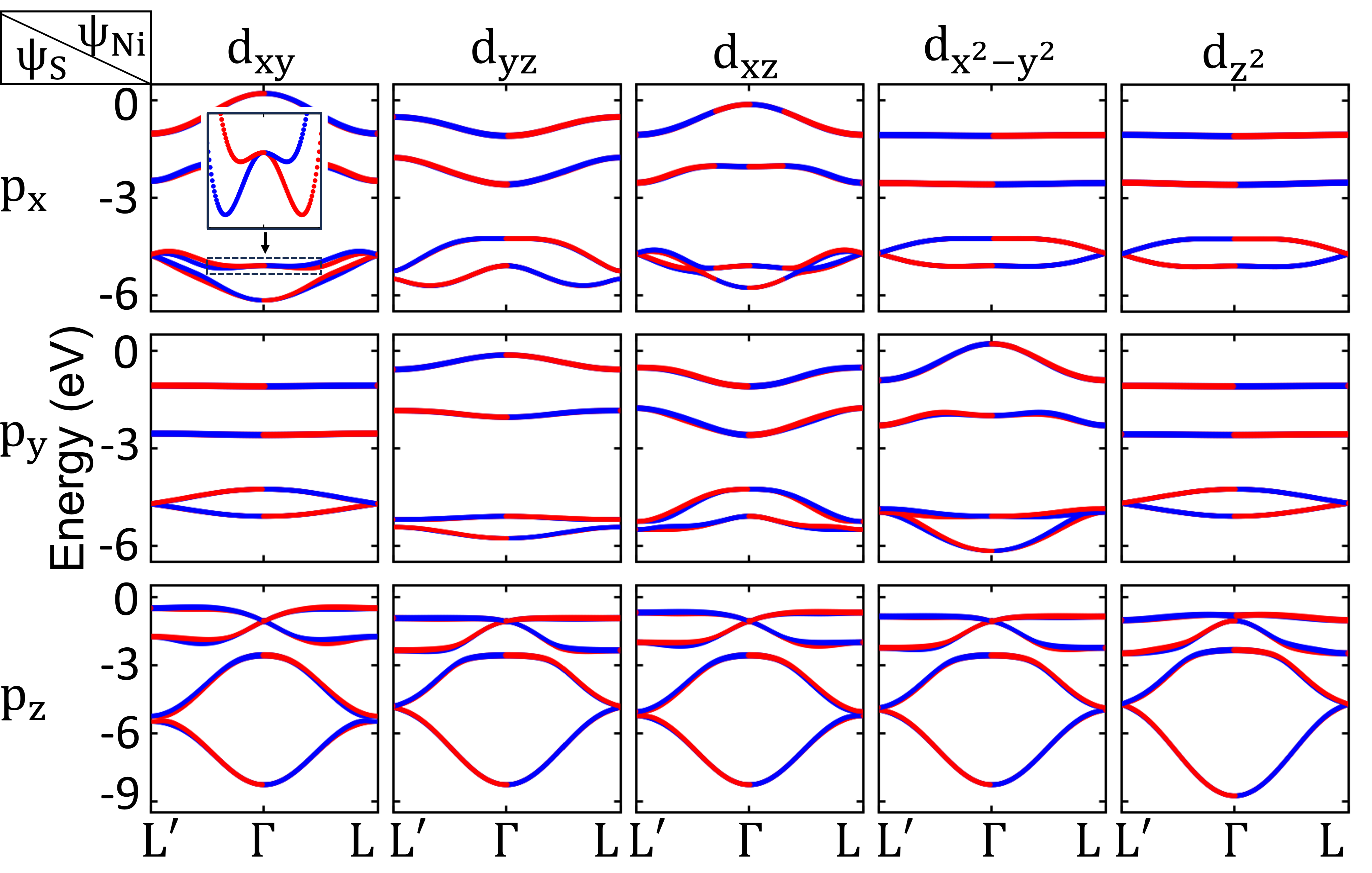}
    \caption{The band structure obtained from the single orbital TB model corresponding to Eq. \ref{upblock1}, where the S-S$^{\prime}$ interaction is present. This S-S$^{\prime}$ interaction is deterministic in the formation of altermagnetism. In the inset, bands are zoomed in to demonstrate the AMSS.}
    \label{im11}
\end{figure}

\textbf{Case II- Single orbital from each of the nickel and sulphur atoms; S-S$^{\boldsymbol{\prime}}$ interaction is present (Fig. \ref{schm}(b)):}
In the previous case, we had blocked the interaction between the inequivalent S atoms S and S$^{\prime}$. Now if we switch on this interaction as schematically shown in Fig. \ref{schm}(b), the Hamiltonian of Eq. \ref{upblock} takes the shape
\begin{eqnarray}
    && \mathcal{H}_{4 \times 4}^{\uparrow \uparrow(\downarrow \downarrow)} = \nonumber \\ 
    && \begin{pmatrix}
        -(+)\frac{\Delta_{\text{Ni}}}{2} + \alpha & 0 & w_1 + iw_2 & w_3 + iw_4\\
        0 & +(-)\frac{\Delta_{\text{Ni}}}{2} + \alpha & v_1 + iv_2 & v_3 + iv_4 \\
        w_1 - iw_2 & v_1 - iv_2 & 0 & u_1 + iu_2 \\
        w_3 - iw_4 & v_3 - iv_4 & u_1 - iu_2 & 0
    \end{pmatrix}.
    \label{upblock1}
\end{eqnarray}
With respect to the characteristic Eq. \ref{eigeneqn}, the new interaction brings the following changes. The coefficients $d_{\uparrow}$ and $d_{\downarrow}$ are not zero but are identical. The coefficients of $\lambda^0$ (i.e., $e_{\uparrow}$ and $e_{\downarrow}$) are no longer identical for spin-up and spin-down blocks.
While the additional term is $+(-) 16 \Delta_{Ni} [ 2u_2 (v_1 v_2 + w_1 w_2) - u_1 (v_1^2 - v_2^2 - w_1^2 + w_2^2)]$ in $e_{\uparrow}$, it is $-(+) 16 \Delta_{Ni} [ 2u_2 (v_1 v_2 + w_1 w_2) - u_1 (v_1^2 - v_2^2 - w_1^2 + w_2^2)]$ in $e_{\downarrow}$ for type I(II) sets of orbitals as defined in table \ref{int-tab6}. Each of these terms indicates that the coupling between Ni and S orbital is modulated by the S-S$^{\prime}$ interaction. Furthermore, the change in the sign suggests that the eigenvalues of spin-up and spin-down blocks are altered equally opposite, and therefore, the antiferromagnetic sublattice spin degeneracy vanishes, which gives rise to AMSS. This is substantiated in Fig. \ref{im11}. Since the additional term in $e_{\uparrow/\downarrow}$ does not contain $\alpha$, one can consider the free atomic onsites energies to be zero to simplify the analysis. Through appendix \ref{perturb-sec}, we show that the AMSS can also be obtained using perturbation theory. Here, the S-S$^{\prime}$ interactions are considered as perturbation to the Hamiltonian described in Eq. \ref{upblock}. Moreover, in appendix \ref{WTB}, we show an alternative approach using Wannier tight binding model (WTB) leads to the same inference.

\begin{figure}[hbt!]
    \centering
   \includegraphics[width=1\linewidth]{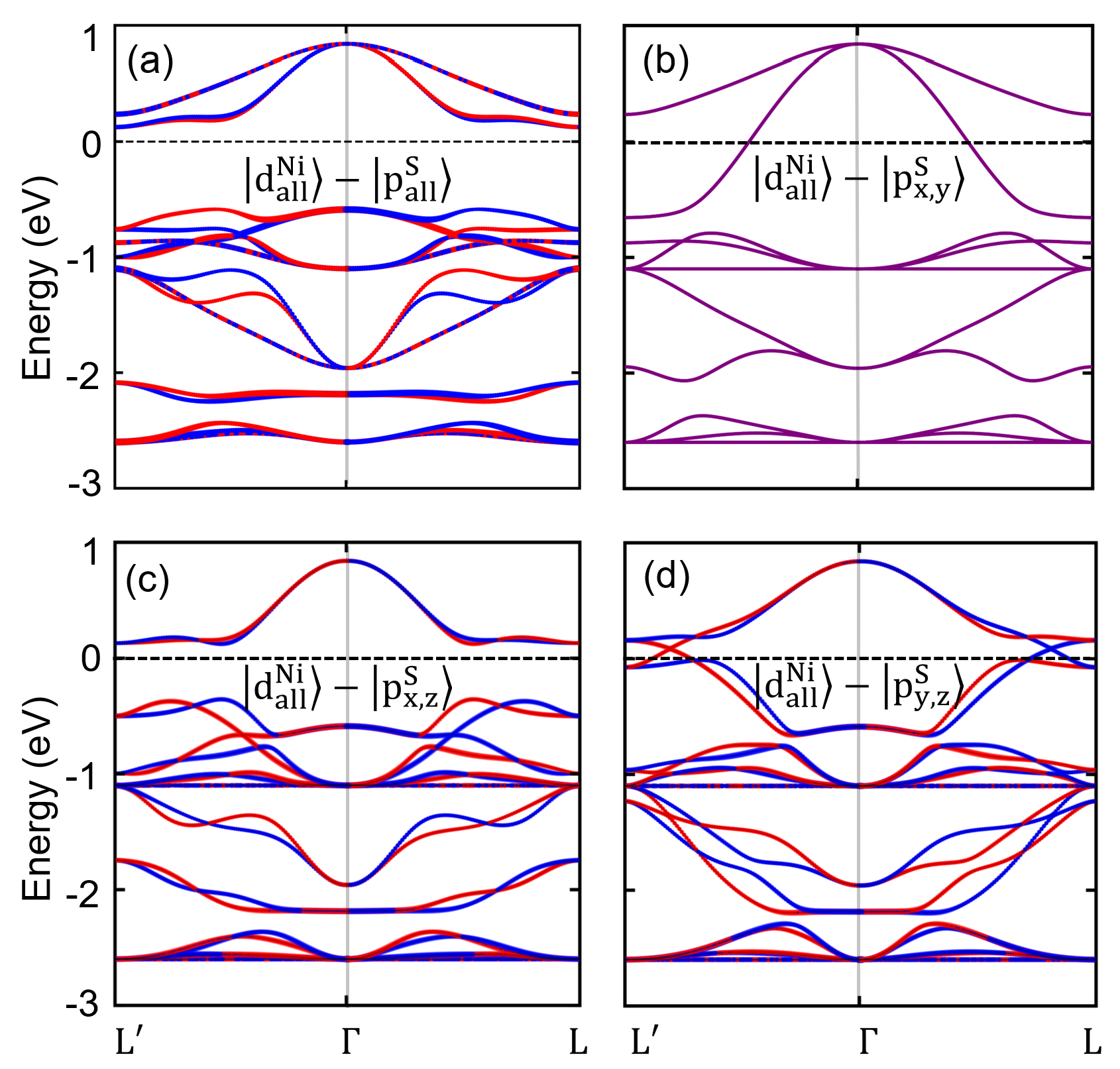}
   \caption{The role of orbital interplay in the formation of altermagnetism in the multi orbital model. Here, only Ni-S interactions are present with antiferromagnetic Hund's coupling. In (a), all Ni-$d$ and S-$p$ orbitals are involved. In (b), (c), and (d), S-$p_z$, S-$p_x$, and S-$p_y$ orbitals are excluded, respectively. In (b), the altermagnetism is absent, while in (c) and (d), the altermagnetism survives.}
    \label{im12}
\end{figure}

\begin{figure}[hbt!]
    \centering
   \includegraphics[width=0.7\linewidth]{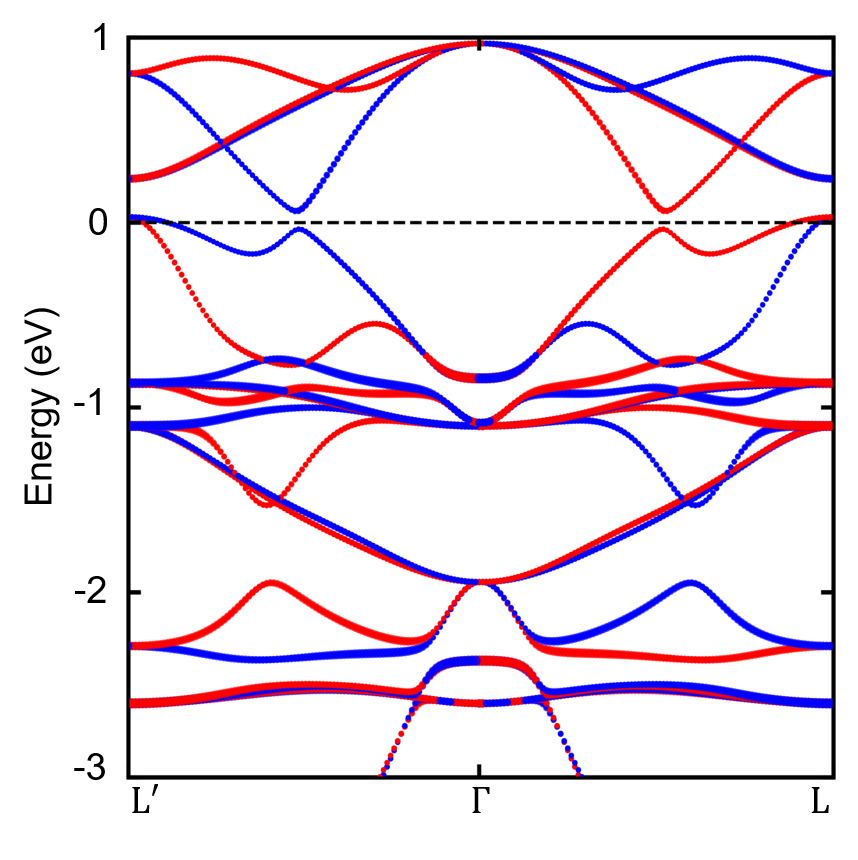}
    \caption{Altermagnetism in the multi orbital model in the presence of all Ni-S interactions, S-S$^{\prime}$ interaction, and the antiferromagnetic Hund's coupling. The S-S$^{\prime}$ interaction amplifies the AMSS width.}
    \label{im13}
\end{figure}

We may note that from the band dispersion point of view, the interaction between the equivalent S atoms and Ni atoms is equally important. However, since they appear in diagonal positions, each of them can be treated as a constant at each k-point. Hence, the nature of the Hamiltonian matrix does not change (see Eqs. \ref{upblock} and \ref{upblock1}) and, therefore, they have no role in producing the AMSS.\\

\textbf{Case III- Multiple orbitals from each of the nickel and sulphur atoms (Fig. \ref{schm}(c) and (d)):} In the above two cases, we presented the necessary conditions for getting altermagnetism in a single orbital model. The important insight we obtain from here is that the momentum-dependent AMSS is possible when individual pair interactions are modulated by the field generated by other interactions. Since in real systems, there are multiple orbitals involved in the interaction, the natural question that arises is whether S-S$^{\prime}$ interaction is necessary. Can the field, due to the interactions from the rest of the $d$- and $p$-orbitals, also introduce AMSS? Therefore, we consider two more situations as schematically illustrated in Fig. \ref{schm}(c) and (d). In (c), all the $d$-orbitals from Ni and all $p$-orbitals from S form the basis. However, the S-S$^{\prime}$ interaction is absent. The resultant band structure is shown in Fig. \ref{im12}, where we find that the altermagnetic features are recovered.
To gain further insight, we examined several subsets of interactions corresponding to Fig. \ref{schm}(c). A few of them are highlighted in Fig. \ref{im12}. We infer that while some subsets exhibit altermagnetism, others do not, which further substantiate the role of orbital symmetries. Specifically, we find that the AMSS appears when a minimum of two $d$-orbitals from Ni sites and a minimum of two $p$-orbitals from S sites, with one being $p_z$, participate in the nearest-neighbor chemical bonding.

Finally, we brought back the S-S$^{\prime}$ interaction into the multi orbital scenario as schematically shown in Fig. \ref{schm}(d). The resultant band structure shows an amplified AMSS, which is of the same order as obtained from the DFT calculations (see Fig. \ref{im13}). The similar observation is also obtained from WTB analyses (see appendix \ref{WTB}).

\section{Summary and Outlook}

In summary, we examined the importance of chemical bonding in the formation of altermagnetism by carrying out DFT and model studies on NiS, a hexagonal NiAs-type crystal. In the past, correlation driven antiferromagnetic metal-insulator transition was demonstrated in this compound. While there was no report on altermagnetism in this compound, a large altermagnetic spin-split (AMSS) is found in the bands of its isostructures such as CrSb and MnTe. From the DFT+U calculations, we find that the competition between localization and enhanced magnetization determine the strength of AMSS. The bottom conduction bands experience a small AMSS for higher $U$ values and for the top valence bands the AMSS increases initially and around for $U=4$ eV, it saturates.

for NiS the AMSS increases with correlation. Most importantly, for the top valence band and bottom conduction bands, the AMSS is greater than 1eV. This makes NiS an ideal playground for experimentally validating altermagnetism and also to explore newer concepts involving it. Furthermore, NiS being an insulator, non-trivial quantum transport can be envisaged out of its altermagnetic valence and conduction band through electron and hole doping.

The crystal chemical bonding, modelled through the Slater-Koster (SK) formalism of linear combination of atomic orbitals, successfully combines crystal symmetries and the orbital symmetries to provide a quantitative description of the underlying mechanisms driving the altermagnetism. Concerning NiS, we find that the deterministic role to induce altermagnetism is played by the second neighbor chemical bonding between the orbitals of the nonmagnetic atoms. These second-neighbor bondings between the inequivalent nonmagnetic atoms S and S$^{\prime}$, modulate the nearest neighbor Ni-S/S$^{\prime}$ and Ni$^{\prime}$-S/S$^{\prime}$ interactions differently. With Ni and Ni$^{\prime}$ forming the opposite spin sublattices, due to this modulation, the antiferromagnetic sublattice band degeneracy is lifted to create AMSS. Single orbital models, where one orbital each from the nickel and sulfur atoms are taken to represent the chemical bonding, create weak altermagnetism. The charge modulation can also occur to induce weak AMSS even if we switch off the  S-S$^{\prime}$ bonding, provided a group of selective orbitals from nickel and sulfur atoms are involved in the nearest neighbor chemical bonding. This highlights the role of orbital symmetries in altermagnetism. We infer that the strong altermagnetism is observed in the NiAs crystal prototype hexagonal family of compounds (NiS, CrSb, MnTe) due to (a) the participation of multiple and selective orbitals, both from the magnetic transition metal atom and the nonmagnetic atoms, and (b) stronger second neighbor interactions among the nonmagnetic atoms. Our study opens up possible pathways to tailor tunable altermagnetism. 

\section*{Acknowledgement}
The authors thank Sashi Satpathy for some useful discussions. AM thanks MoE India for the PMRF fellowship.

\section*{Data Availibility}
The data that support the findings of this article are not publicly available. The data are available from the authors upon reasonable request.

%%\newpage
\appendix
\section{The nonmagnetic electronic structure}
\label{nonmagnetic}
\begin{figure}[hbt!]
    \centering
    \includegraphics[width=1.0\linewidth]{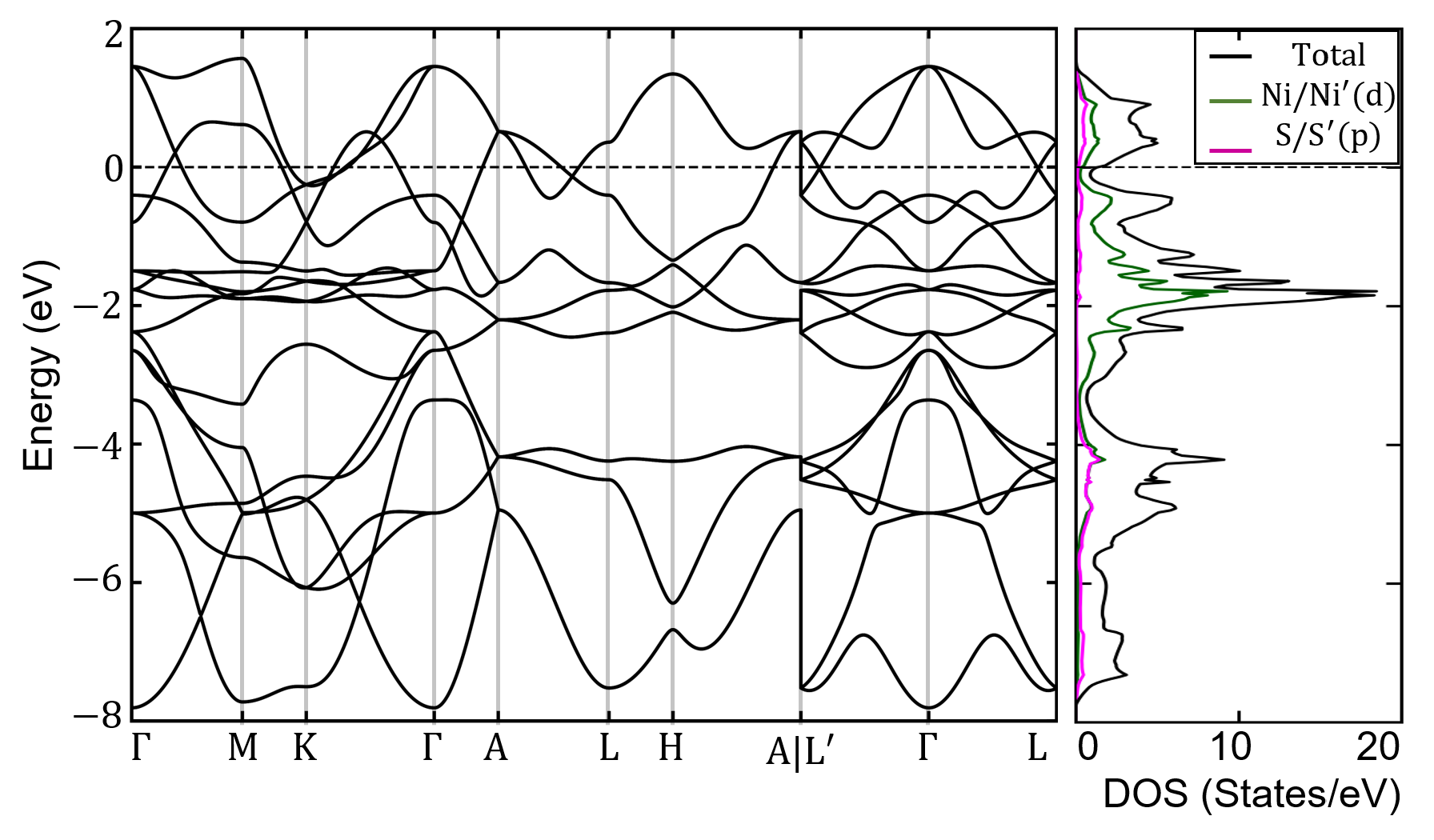}
    \caption{The nonmagnetic bandstructure (left) and the DOS (right) of NiS are shown for $U = 0$. The Fermi level is shown by black dotted line.}
    \label{NM_U=0}
\end{figure}

As discussed in the main text, NiS stabilizes in a nonmagnetic ground state for very low onsite Coulomb repulsion ($< 1$ eV). The nonmagnetic band structure and DOS for $U = 0$ are shown in Fig. \ref{NM_U=0}. Depleted DOS at the $E_F$ resembles a pseudo gap, which suggests that it is nearly semimetallic even in the nonmagnetic phase.
%%%%%%%%%%%%%%%%%%%%%%%%%%%%%%%%%%%%%%%

\section{Direction cosines ($l, m, n$) and Hopping interactions}
\label{hopping}
In the SK formalism of the TB model, within the LCAO framework, the hopping integrals between a pair of orbitals are expressed through the bonding parameters ($\sigma$, $\pi$, $\delta$) and the direction cosines ($l,\ m,\ n$). The latter have a central role as they are determined through the position of the atoms and hence are intertwined with the crystal symmetry. Below, we list the values of $l,\ m$, and $n$ for several atom pairs restricted to the second nearest neighbor both in the plane and out of the plane. This will help in identifying the allowed and forbidden chemical bondings among the valence orbitals in NiS (see table \ref{int-tab2}, \ref{int-tab3}, \ref{int-tab4}, and \ref{int-tab5}). 
 
 Below, we explicitly derive the interaction between Ni$-d_{xy}$ - S$-p_x$ (represented by the term $w_1 + iw_2$ (Eq. \ref{supp-eq1})), Ni$^{\prime}-d_{xy}$ - S$-p_x$ (represented by the term $v_1 + iv_2$ \ref{supp-eq2}), and S$-p_x$ - S$^{\prime}-p_x$ (represented by the term $u_1 + iu_2$ \ref{supp-eq3}) orbitals as an example. Here, we neglect the $\delta$ interaction as it is very small in magnitude. All the other interactions can be written by following the same prescription.

\begin{widetext}
    \begin{eqnarray}
        \langle Ni-d_{xy}|  H  | S-p_x\rangle &=& w_1 + i w_2 = \sum_{i=1}^{3} (\sqrt{3}l_i^2m_i t_{ns\sigma} + m_i(1-l_i^2)t_{ns\pi}) e^{i\boldsymbol{k}\cdot \boldsymbol{R}_{ij}}~~~~~~~~~~~\text{(ref: table \ref{int-tab2})}\nonumber \\
        &=& 2\left(t_{ns\sigma}\frac{a^3}{8A^3} + t_{ns\pi} \frac{a}{2\sqrt{3}A}\left(1-\frac{a^2}{4B^2}\right)\right)\cos{\left(\frac{a}{2}k_x\right)}e^{i(\frac{a}{2\sqrt{3}}k_y-\frac{c}{4}k_z)}- t_{ns\pi} \frac{a}{\sqrt{3}A}e^{i(-\frac{a}{\sqrt{3}}k_y-\frac{c}{4}k_z)}\nonumber \\
    \label{supp-eq1}
    \end{eqnarray}

    \begin{eqnarray}
        \langle Ni'-d_{xy}|  H  | S-p_x\rangle &=& v_1 + i v_2 = \sum_{i=1}^{3} (\sqrt{3}l_i^2m_i t_{ns\sigma} + m_i(1-l_i^2)t_{ns\pi}) e^{i\boldsymbol{k}\cdot \boldsymbol{R}_{ij}}~~~~~~~~~~~\text{(ref: table \ref{int-tab2})}\nonumber \\
        &=& 2\left(t_{ns\sigma}\frac{a^3}{8A^3} + t_{ns\pi} \frac{a}{2\sqrt{3}A}\left(1-\frac{a^2}{4B^2}\right)\right)\cos{\left(\frac{a}{2}k_x \right)}e^{i(\frac{a}{2\sqrt{3}}k_y+\frac{c}{4}k_z)}-t_{ns\pi} \frac{a}{\sqrt{3}A}e^{i(-\frac{a}{\sqrt{3}}k_y+\frac{c}{4}k_z)}\nonumber \\
    \label{supp-eq2}
    \end{eqnarray}

    \begin{eqnarray}
        \langle S-p_{x}| H| S'-p_x\rangle &=& u_1 + i u_2 = \sum_{i=1}^{6} (l_i^2 t_{s\sigma 2} +(1-l_i^2) t_{s\pi 2}) e^{i\boldsymbol{k}\cdot \boldsymbol{R}_{ij}}~~~~~~~~~~~~~~~~~~~~~~~~~~~~~~~~~~~~~\text{(ref: table \ref{int-tab3})}\nonumber \\
        &=& 4\left(t_{s\sigma 2} \frac{a^2}{4B^2}-t_{s\pi 2}\left(1-\frac{a^2}{4B^2}\right)\right)\cos{\left(\frac{a}{2}k_x\right)}\cos{\left(\frac{c}{2}k_z\right)}e^{i\frac{a}{2\sqrt{3}}k_y} + 2t_{s\pi 2}\cos{(\frac{c}{2}k_z)}e^{-i\frac{a}{\sqrt{3}}k_y})\nonumber \\
    \label{supp-eq3}
    \end{eqnarray}
\end{widetext}

\begin{table}[hbt!]
\begin{center}
\caption{The ($l,\ m,\ n$) of the sulfur atoms with nickel atom at the centre. Here, Ni and Ni$^{\prime}$ represent the two inequivalent Ni atoms. They also form the two antiferromagnetic sublattices. Here, $A=\sqrt{\frac{a^2}{3}+\frac{c^2}{16}}$}
\setlength{\tabcolsep}{6pt}
\renewcommand{\arraystretch}{1.5}
\begin{tabular}{c c| c c}
\hline
\hline
 & Ni & Ni$^{\prime}$ \\
\hline
S$_1$ & $\frac{1}{A}(\frac{a}{2},\frac{a}{2\sqrt{3}},-\frac{c}{4})$ & $\frac{1}{A}(\frac{a}{2},\frac{a}{2\sqrt{3}},\frac{c}{4})$ \\
S$_2$ & $\frac{1}{A}(-\frac{a}{2},\frac{a}{2\sqrt{3}},-\frac{c}{4})$ & $\frac{1}{A}(-\frac{a}{2},\frac{a}{2\sqrt{3}},\frac{c}{4})$ \\
S$_3$ & $\frac{1}{A}(0,-\frac{a}{\sqrt{3}},-\frac{c}{4})$ & $\frac{1}{A}(0,-\frac{a}{\sqrt{3}},\frac{c}{4})$ \\
S$^{\prime}_1$ & $\frac{1}{A}(\frac{a}{2},-\frac{a}{2\sqrt{3}},\frac{c}{4})$ & $\frac{1}{A}(\frac{a}{2},-\frac{a}{2\sqrt{3}},-\frac{c}{4})$ \\
S$^{\prime}_2$ & $\frac{1}{A}(0,\frac{a}{\sqrt{3}},\frac{c}{4})$ & $\frac{1}{A}(0,\frac{a}{\sqrt{3}},-\frac{c}{4})$ \\
S$^{\prime}_3$ & $\frac{1}{A}(-\frac{a}{2},-\frac{a}{2\sqrt{3}},\frac{c}{4})$ & $\frac{1}{A}(-\frac{a}{2},-\frac{a}{2\sqrt{3}},-\frac{c}{4})$ \\
\hline
\hline
\end{tabular}
\label{int-tab2}
\end{center}
\end{table}

\begin{table}[hbt!]
\begin{center}
\caption{The ($l,\ m,\ n$) of the S$^{\prime}$ atoms with S atom at the center. It may be noted that, like the case of nickel atoms, there are two inequivalent sulfur atoms. Here, $B=\sqrt{\frac{a^2}{3}+\frac{c^2}{4}}$}
\setlength{\tabcolsep}{6pt}
\renewcommand{\arraystretch}{1.5}
\begin{tabular}{c| c c}
\hline
\hline
 & Upper layer & Lower layer \\
\hline
S$^{\prime}_1$ & $\frac{1}{B}(-\frac{a}{2},\frac{a}{2\sqrt{3}},\frac{c}{2})$ & $\frac{1}{B}(-\frac{a}{2},\frac{a}{2\sqrt{3}},-\frac{c}{2})$ \\
S$^{\prime}_2$ & $\frac{1}{B}(\frac{a}{2},\frac{a}{2\sqrt{3}},\frac{c}{2})$ & $\frac{1}{B}(\frac{a}{2},\frac{a}{2\sqrt{3}},-\frac{c}{2})$ \\
S$^{\prime}_3$ & $\frac{1}{B}(0,-\frac{a}{\sqrt{3}},\frac{c}{2})$ & $\frac{1}{B}(0,-\frac{a}{\sqrt{3}},-\frac{c}{2})$ \\
\hline
\hline
\end{tabular}
\label{int-tab3}
\end{center}
\end{table}

\begin{table}[hbt!]
\begin{center}
\caption{The $(l,\ m,\ n)$ of the neighboring Ni$^{\prime}$ atoms for a given Ni atom at the center. Here, $D=\sqrt{a^2+\frac{c^2}{4}}$ }
\setlength{\tabcolsep}{6pt}
\renewcommand{\arraystretch}{1.5}
\begin{tabular}{c| c c}
\hline
\hline
 & Upper layer & Lower layer \\
\hline
Ni$^{\prime}_1$ & $(0,0,1)$ & $(0,0,-1)$ \\
Ni$^{\prime}_2$ & $\frac{1}{D}(a,0,\frac{c}{2})$ & $\frac{1}{D}(a,0,-\frac{c}{2})$ \\
Ni$^{\prime}_3$ & $\frac{1}{D}(-a,0,\frac{c}{2})$ & $\frac{1}{D}(-a,0,-\frac{c}{2})$ \\
Ni$^{\prime}_4$ & $\frac{1}{D}(\frac{a}{2},\frac{\sqrt{3}}{2}a,\frac{c}{2})$ & $\frac{1}{D}(\frac{a}{2},\frac{\sqrt{3}}{2}a,-\frac{c}{2})$ \\
Ni$^{\prime}_5$ & $\frac{1}{D}(-\frac{a}{2},\frac{\sqrt{3}}{2}a,\frac{c}{2})$ & $\frac{1}{D}(-\frac{a}{2},\frac{\sqrt{3}}{2}a,-\frac{c}{2})$ \\
Ni$^{\prime}_6$ & $\frac{1}{D}(\frac{a}{2},-\frac{\sqrt{3}}{2}a,\frac{c}{2})$ & $\frac{1}{D}(\frac{a}{2},-\frac{\sqrt{3}}{2}a,-\frac{c}{2})$ \\
Ni$^{\prime}_7$ & $\frac{1}{D}(-\frac{a}{2},-\frac{\sqrt{3}}{2}a,\frac{c}{2})$ & $\frac{1}{D}(-\frac{a}{2},-\frac{\sqrt{3}}{2}a,-\frac{c}{2})$ \\
\hline
\hline
\end{tabular}
\label{int-tab4}
\end{center}
\end{table}

\begin{table}[hbt!]
\begin{center}
\caption{The $(l,\ m,\ n)$ of the neighboring Ni and S atoms of one kind with a given Ni or S atom of the same kind at the centre. This helps in expressing the hopping integrals for the Ni-Ni, Ni$^{\prime}$-Ni$^{\prime}$, S-S, and S$^{\prime}$-S$^{\prime}$  interactions.}
\setlength{\tabcolsep}{6pt}
\renewcommand{\arraystretch}{1.5}
\begin{tabular}{c| c}
\hline
\hline
 & Ni/Ni$^{\prime}$/S/S$^{\prime}$ \\
neighboring atoms & reference centre \\
\hline
Ni$_1$/Ni$^{\prime}_1$/S$_1$/S$^{\prime}_1$ & $(1,0,0)$ \\
Ni$_2$/Ni$^{\prime}_2$/S$_2$/S$^{\prime}_2$& $(-1,0,0)$ \\
Ni$_3$/Ni$^{\prime}_3$/S$_3$/S$^{\prime}_3$& $(\frac{1}{2}, -\frac{\sqrt{3}}{2}, 0)$  \\
Ni$_1$/Ni$^{\prime}_4$/S$_4$/S$^{\prime}_4$ & $(-\frac{1}{2}, -\frac{\sqrt{3}}{2}, 0)$  \\
Ni$_5$/Ni$^{\prime}_5$/S$_5$/S$^{\prime}_5$ & $(\frac{1}{2}, \frac{\sqrt{3}}{2}, 0)$  \\
Ni$_6$/Ni$^{\prime}_6$/S$_6$/S$^{\prime}_6$ & $(-\frac{1}{2}, \frac{\sqrt{3}}{2}, 0)$  \\
\hline
\hline
\end{tabular}
\label{int-tab5}
\end{center}
\end{table}

Using the aforementioned equations, one can construct the single orbital model Hamiltonians for each spin block as given in Eq. \ref{upblock1} in the main text. The characteristic equations for each matrix are shown in Eq. \ref{eigeneqn}. The coefficients of the characteristic equation for the spin-up block are given by,
\begin{eqnarray}
    %b_{\uparrow} &=& 0 
    b_{\uparrow} &=& -4 \alpha
    \label{coeff-eq1}
\end{eqnarray}

\begin{eqnarray}
    c_{\uparrow} &=& -4 (u_1^2 + u_2^2 +  v_1^2 + v_2^2 \nonumber \\ 
    &+&  v_3^2 +  v_4^2 +  w_1^2 +  w_2^2 +  w_3^2 +  w_4^2 -\alpha^2) - \Delta^2
    \label{coeff-eq2}
\end{eqnarray}

\begin{eqnarray}
    d_{\uparrow} &=& - 16 (u_1 v_1 v_3 -  u_2 v_2 v_3 +  u_2 v_1 v_4 +  u_1 v_2 v_4 \nonumber \\ 
    &+& u_1 w_1 w_3 - u_2 w_2 w_3 + u_2 w_1 w_4 + u_1 w_2 w_4) \nonumber \\
    &+& 8 \alpha (2 u_1^2 + 2 u_2^2 + v_1^2 + v_2^2 + v_3^2 + v_4^2 \nonumber \\
   &+& w_1^2 + w_2^2 + w_3^2 + 
   w_4^2) \nonumber \\
    &+& 4 \Delta (v_1^2 + v_2^2 + v_3^2 + v_4^2 - w_1^2 - w_2^2 - w_3^2 - w_4^2) \nonumber \\
    \label{coeff-eq3}
\end{eqnarray}

\begin{eqnarray}
    e_{\uparrow} &=& 16 (v_3^2 w_1^2 + v_4^2 w_1^2 + v_3^2 w_2^2 + v_4^2 w_2^2 \nonumber \\
    &+& v_1^2 w_3^2 + v_2^2 w_3^2 + v_1^2 w_4^2 + v_2^2 w_4^2) \nonumber \\
    &+& 32 (- v_1 v_3 w_1 w_3 - v_2 v_4 w_1 w_3 - v_2 v_3 w_2 w_3 + v_1 v_4 w_2 w_3 \nonumber \\
    &+& v_2 v_3 w_1 w_4 - v_1 v_4 w_1 w_4 -  v_1 v_3 w_2 w_4 - v_2 v_4 w_2 w_4 ) \nonumber \\
    &+& 32 \alpha (32 u_1 v_1 v_3 - u_2 v_2 v_3 + u_2 v_1 v_4 + u_1 v_2 v_4 \nonumber \\
    &+& u_1 w_1 w_3 - u_2 w_2 w_3 + u_2 w_1 w_4 + u_1 w_2 w_4) \nonumber \\
    &-& 16 \Delta (u_1 v_1 v_3 - u_2 v_2 v_3 + u_2 v_1 v_4 + u_1 v_2 v_4  \nonumber \\
    &-& u_1 w_1 w_3 + u_2 w_2 w_3 - u_2 w_1 w_4 - u_1 w_2 w_4) \nonumber \\
    &+& 4 \Delta^2 (u_1^2 + u_2^2) -16 \alpha^2 (u_1^2 + u_2^2)
    \label{coeff-eq4}
\end{eqnarray}

The same for the spin-down block is written as,

\begin{eqnarray}
    %b_{\downarrow} &=& 0
    b_{\downarrow} &=& -4\alpha
    \label{coeff-eq5}
\end{eqnarray}

\begin{eqnarray}
    c_{\downarrow} &=& -4 (u_1^2 + u_2^2 +  v_1^2 + v_2^2 \nonumber \\ 
    &+&  v_3^2 +  v_4^2 +  w_1^2 +  w_2^2 +  w_3^2 +  w_4^2 -\alpha^2) - \Delta^2
    \label{coeff-eq6}
\end{eqnarray}

\begin{eqnarray}
    d_{\downarrow} &=& - 16 (u_1 v_1 v_3 -  u_2 v_2 v_3 +  u_2 v_1 v_4 +  u_1 v_2 v_4 \nonumber \\ 
    &+& u_1 w_1 w_3 - u_2 w_2 w_3 + u_2 w_1 w_4 + u_1 w_2 w_4) \nonumber \\
    &+& 8 \alpha (2 u_1^2 + 2 u_2^2 + v_1^2 + v_2^2 + v_3^2 + v_4^2 \nonumber \\
   &+& w_1^2 + w_2^2 + w_3^2 + 
   w_4^2) \nonumber \\
    &-& 4 \Delta (v_1^2 + v_2^2 + v_3^2 + v_4^2 - w_1^2 - w_2^2 - w_3^2 - w_4^2) \nonumber \\
    \label{coeff-eq7}
\end{eqnarray}

\begin{eqnarray}
    e_{\downarrow} &=& 16 (v_3^2 w_1^2 + v_4^2 w_1^2 + v_3^2 w_2^2 + v_4^2 w_2^2 \nonumber \\
    &+& v_1^2 w_3^2 + v_2^2 w_3^2 + v_1^2 w_4^2 + v_2^2 w_4^2) \nonumber \\
    &+& 32 (- v_1 v_3 w_1 w_3 - v_2 v_4 w_1 w_3 - v_2 v_3 w_2 w_3 + v_1 v_4 w_2 w_3 \nonumber \\
    &+& v_2 v_3 w_1 w_4 - v_1 v_4 w_1 w_4 -  v_1 v_3 w_2 w_4 - v_2 v_4 w_2 w_4 ) \nonumber \\
    &+& 32 \alpha (32 u_1 v_1 v_3 - u_2 v_2 v_3 + u_2 v_1 v_4 + u_1 v_2 v_4 \nonumber \\
    &+& u_1 w_1 w_3 - u_2 w_2 w_3 + u_2 w_1 w_4 + u_1 w_2 w_4) \nonumber \\
    &+& 16 \Delta (u_1 v_1 v_3 - u_2 v_2 v_3 + u_2 v_1 v_4 + u_1 v_2 v_4  \nonumber \\
    &-& u_1 w_1 w_3 + u_2 w_2 w_3 - u_2 w_1 w_4 - u_1 w_2 w_4) \nonumber \\
    &+& 4 \Delta^2 (u_1^2 + u_2^2) -16 \alpha^2 (u_1^2 + u_2^2)
    \label{coeff-eq8}
\end{eqnarray}
From these equations, the difference in the coefficient of $\lambda^0$ for spin-up and spin-down blocks is computed by imposing the conditions of table \ref{int-tab6}. This, in turn, gives rise to altermagnetism.

\section{Realization of altermagnetism using perturbation theory}
\label{perturb-sec}
\begin{figure}[hbt!]
    \centering
    \includegraphics[width=1\linewidth]{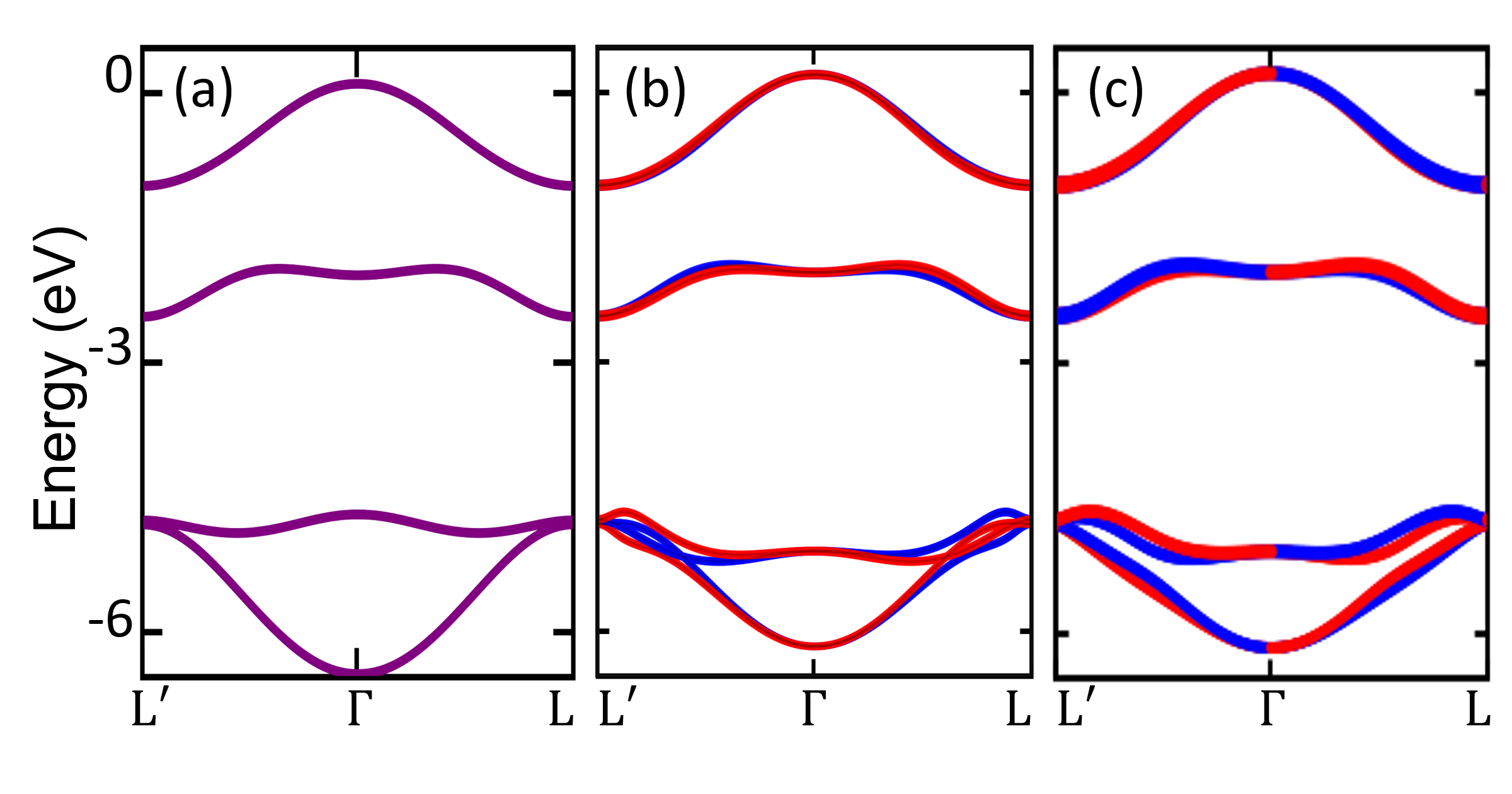}
    \caption{Formation of altermagnet from antiferromagnet via perturbation. (a) The band structure with nearest neighbor Ni-S and antiferromagnetic Hund's coupling interactions. This produces a pure antiferromagnetic band structure. (b) The band structure of (a) is perturbed by S-S$^{\prime}$ interactions, which produces AMSS. (c) The band structure obtained from the single orbital minimal Hamiltonian as described in case-II of section \ref{TB-sec}.} 
    \label{perturbation}
\end{figure}
To get further insight into the role of S-S$^{\prime}$ interaction in the formation of altermagnetism, we do a perturbation theory analysis for the single orbital model. We first consider the Hamiltonian of Case-I of section \ref{TB-sec} that contains the nearest neighbor Ni-S interactions along with antiferromagnetic Hund's coupling interaction. This Hamiltonian is represented as
% as depicted in Case-I of section \ref{TB-sec}. To understand the band structure further, we rewrite the Hamiltonian $\mathcal{H}^0$ (see Eq. \ref{upblock}) with the following basis order $|\text{Ni}-d \rangle$, $|\text{Ni}^{\prime}-d \rangle$, $|\text{S}-p \rangle$, $|\text{S}^{\prime}-p \rangle$ for up spin block. The spin-up block of the new Hamiltonian takes the form,

\begin{equation}
    \mathcal{H}_{4 \times 4}^{0\uparrow \uparrow} = 
    \begin{pmatrix}
        -\frac{\Delta_{\text{Ni}}}{2} & 0 & w_1 + iw_2 & w_3 + iw_4\\
        0 & +\frac{\Delta_{\text{Ni}}}{2} & v_1 + iv_2 & v_3 + iv_4 \\
        w_1 - iw_2 & v_1 - iv_2 & 0 & 0 \\
        w_3 - iw_4 & v_3 - iv_4 & 0 & 0
    \end{pmatrix}.
    \label{upblock2}
\end{equation}
Here, we consider $\alpha=0$ as it has no role in the formation of AMSS. This Hamiltonian can be divided into four sub-blocks, which can be written as,
\begin{equation}
    \mathcal{H}_{4 \times 4}^{0\uparrow \uparrow} = 
    \begin{pmatrix}
        A_{2 \times 2}^{0\uparrow \uparrow} & B_{2 \times 2}^{0\uparrow \uparrow}\\
        C_{2 \times 2}^{0\uparrow \uparrow} & D_{2 \times 2}^{0\uparrow \uparrow} 
    \end{pmatrix},
    \label{upblock3}
\end{equation}
where, 
\begin{eqnarray}
    A_{2 \times 2}^{0\uparrow \uparrow} &=& \begin{pmatrix}
        -\frac{\Delta_{\text{Ni}}}{2} & 0\\
        0 & +\frac{\Delta_{\text{Ni}}}{2} 
    \end{pmatrix}, \\
    B_{2 \times 2}^{0\uparrow \uparrow} &=& \begin{pmatrix}
        w_1 + iw_2 & w_3 + iw_4\\
        v_1 + iv_2 & v_3 + iv_4 
    \end{pmatrix},\\
    C_{2 \times 2}^{0\uparrow \uparrow} &=& \begin{pmatrix}
        w_1 - iw_2 & v_1 - iv_2\\
        w_3 - iw_4 & v_3 - iv_4 
    \end{pmatrix}, \\
    D_{2 \times 2}^{0\uparrow \uparrow} &=& \begin{pmatrix}
        0 & 0\\
        0 & 0
    \end{pmatrix}.
\end{eqnarray}
The characteristic equation can be written as,
\begin{eqnarray}
    |\mathcal{H}_{4 \times 4}^{0\uparrow \uparrow} - \lambda I_{4 \times 4}| &=& 0 \\
    \begin{vmatrix}
        A_{2 \times 2}^{\uparrow \uparrow} -\lambda I_{2\times 2} & B_{2 \times 2}^{\uparrow \uparrow}\\
        C_{2 \times 2}^{\uparrow \uparrow} & D_{2 \times 2}^{\uparrow \uparrow} -\lambda I_{2\times 2}
    \end{vmatrix} &=& 0
    \label{upblock4}
\end{eqnarray}
Since $C_{2 \times 2}^{\uparrow \uparrow}$ and $(D_{2 \times 2}^{\uparrow \uparrow} -\lambda I_{2\times 2})$ are square matrices of the same size and they commute with each other here, Eq.\ref{upblock4} can also be represented as,
\begin{equation}
    |(A_{2 \times 2}^{\uparrow \uparrow} -\lambda I_{2\times 2})(D_{2 \times 2}^{\uparrow \uparrow} -\lambda I_{2\times 2}) - B_{2 \times 2}^{\uparrow \uparrow}C_{2 \times 2}^{\uparrow \uparrow}| = 0
\end{equation}

The characteristic equation for the spin-down block can be obtained by following the same procedure. In the spin-down block, all the subblocks are the same as that of the spin-up block except $A_{2 \times 2}^{0\downarrow \downarrow}$ which is given by,
\begin{eqnarray}
    A_{2 \times 2}^{0\downarrow \downarrow} &=& \begin{pmatrix}
        \frac{\Delta_{\text{Ni}}}{2} & 0\\
        0 & -\frac{\Delta_{\text{Ni}}}{2} 
    \end{pmatrix}.
\end{eqnarray}
Therefore, the characteristic equations of the spin-up and spin-down blocks are identical, ensuring ideal antiferromagnetic band structure (i.e., $\epsilon_{\uparrow}^{SL-1}(k)$ = $\epsilon_{\downarrow}^{SL-2}(k)$ = $\epsilon_{\downarrow}^{SL-2}(-k)$). Hence, it is proper to consider that $\mathcal{H}_{4\times 4}^{0\uparrow \uparrow}$ and $\mathcal{H}_{4\times 4}^{0\downarrow \downarrow}$ form the unperturbed Hamiltonian. 
The S-S$^{\prime}$ second neighbor interactions, appearing in the subblocks $D_{2 \times 2}^{0\uparrow \uparrow}$ and $D_{2 \times 2}^{0\downarrow \downarrow}$, act as perturbation, i.e., $\mathcal{H}_{(4 \times 4)} = \mathcal{H}_{(4 \times 4)}^{0\uparrow \uparrow(\downarrow \downarrow)} + \mathcal{H}_{(4 \times 4)}^{p\downarrow \downarrow)}$ where,

\begin{figure}[b]
    \centering
    \includegraphics[width=1.0\linewidth]{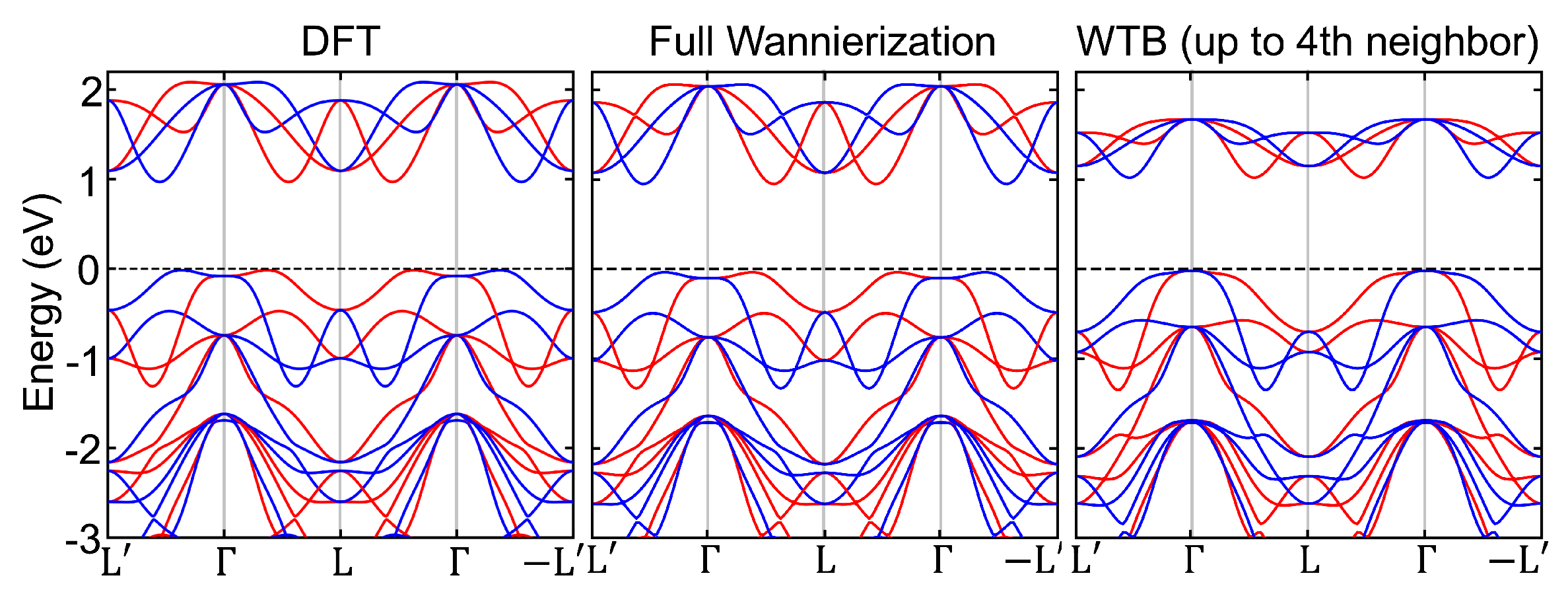}
    \caption{The comparison of the Wannier bands with that of the DFT. The leftmost figure represents the DFT obtained band structure with $U = 4$ eV, while the middle figure is the Wannier band structure with full Wannierization. There is a perfect match between DFT bands and Wannier bands. The rightmost figure represents the band structure obtained from WTB, where the interactions are considered up to a length scale of $4.4\ \AA$ (i.e., up to 4th nearest neighbor)}
    \label{dft-wann}
\end{figure}

\begin{eqnarray}
    \mathcal{H}_{4 \times 4}^{p\uparrow \uparrow (\downarrow \downarrow)} = 
    \begin{pmatrix}
        0 & 0\\
        0 & D_{2 \times 2}^{p\uparrow \uparrow(\downarrow \downarrow)} 
    \end{pmatrix},
    \label{perturb2}
\end{eqnarray}
% In $\mathcal{H}_{8 \times 8}^{p}$, all the $2\times 2$ subblock matrices are null matrix except
\begin{equation*}
    D_{2 \times 2}^{p\uparrow \uparrow} = \begin{pmatrix}
        0 & u_1 + iu_2\\
        u_1 - iu_2 & 0
    \end{pmatrix}
\end{equation*}
The eigenvalues of $\mathcal{H}$ can be computed from $\mathcal{H}^0$ with first order perturbation correction as,
\begin{equation}
    \epsilon_n = \epsilon_n^0 + \langle \phi_n | \mathcal{H}^p | \phi_n \rangle
\end{equation}
where $|\phi_n \rangle$s are the eigenstates of the unperturbed Hamiltonian. The band structure is shown in Fig. \ref{perturbation}. From the figure, one can see that band structures obtained from the original Hamiltonian and the perturbation are in good agreement with each other, which reveals that the S-S$^{\prime}$ interactions perturb the antiferromagnetic band structure and form the altermagnetic state.

\section{Results from the Wannier tight binding model}
\label{WTB}

\begin{figure}[hbt!]
    \centering
    \includegraphics[width=1\linewidth]{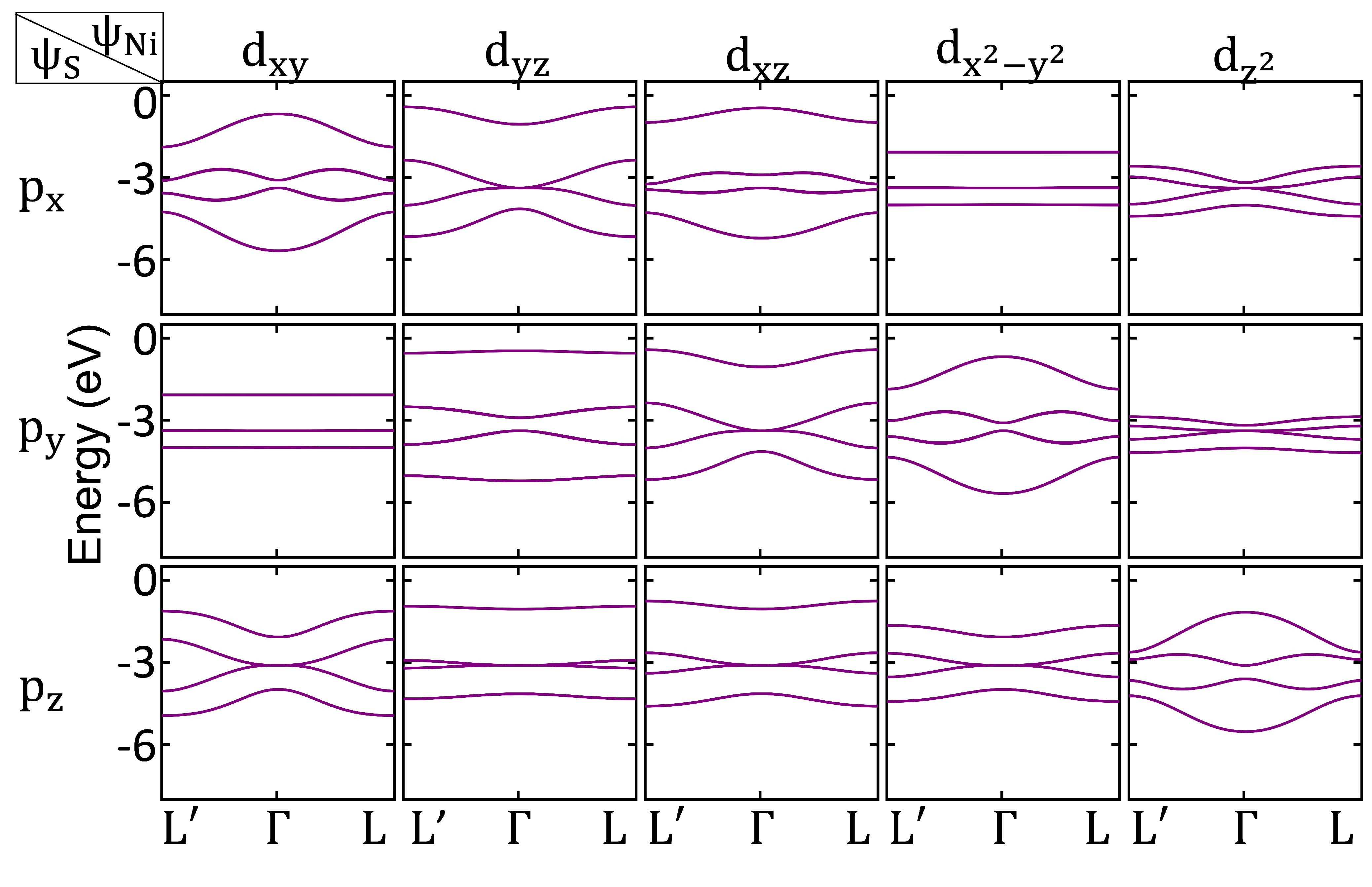}
    \caption{WTB band structure for single orbital model (i.e, one $d$ orbital from Ni and one $p$ orbital from S) in the presence of nearest neighbor Ni-S interaction and Hund's coupling. This indicates that the aforementioned interactions are insufficient to trigger altermagnetism}
    \label{single-orb1}
\end{figure}

\begin{figure}[hbt!]
    \centering
    \includegraphics[width=1\linewidth]{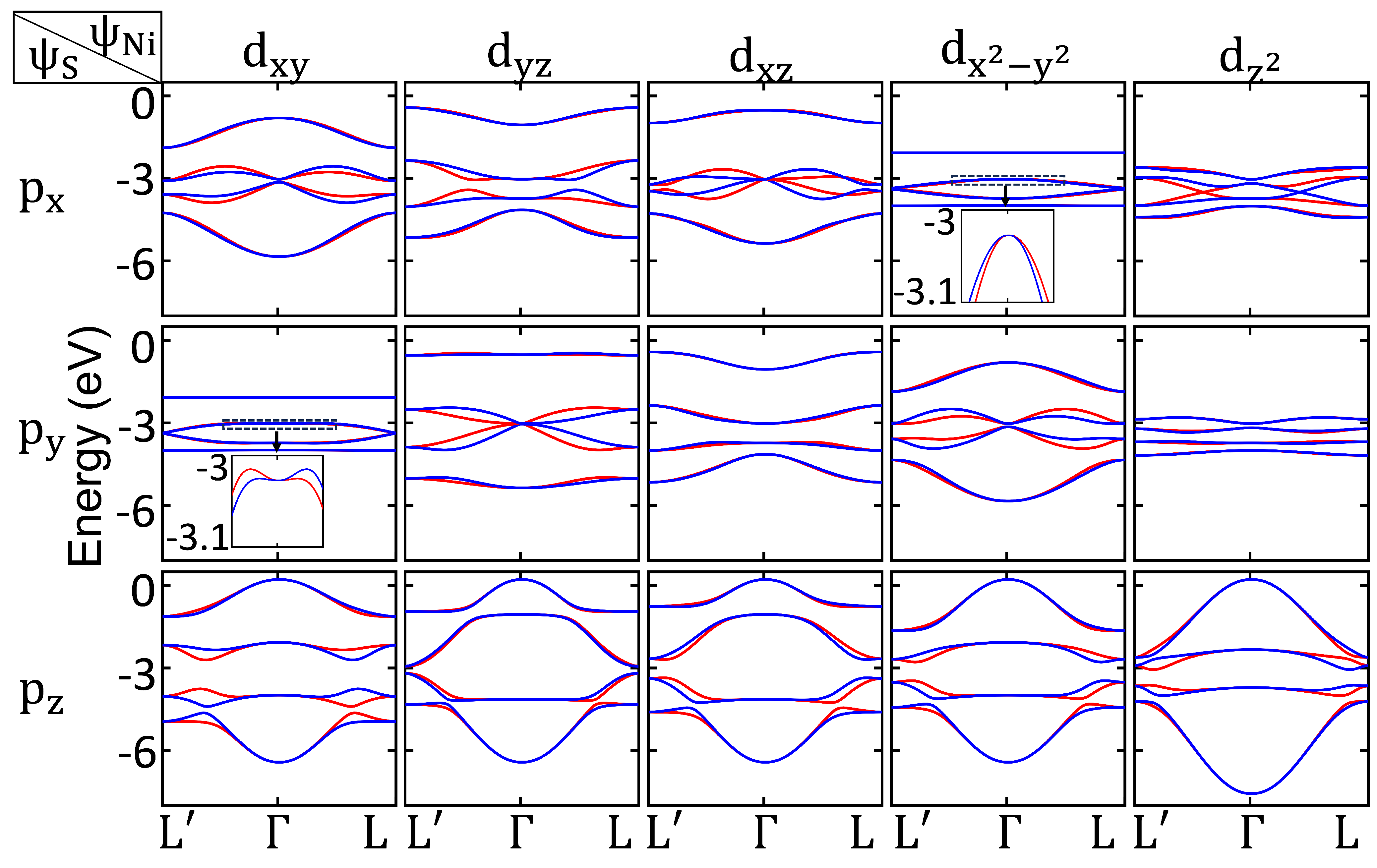}
    \caption{Single orbital Wannier-TB model with Hund's coupling, all nearest neighbor Ni-S interaction, S-S$^{\prime}$ interactions. Here, we see that S-S$^{\prime}$ interaction triggers the altermagnetism. In the inset the zoomed version of two AMSS are shown for better visualization}
    \label{single-orb2}
\end{figure}

The alternate way of constructing a tight binding model is through the Wannierization where the maximally localized Wannier functions (MLWFs) are treated as the bases of the Hamiltonian. In this Appendix, we present the band structure obtained from WTB using WANNIER90 program \cite{wannier} to further substantiate the role of interactions among the orbitals of the nonmagnetic atoms in inducing the altermagnetism.

\begin{figure}[hbt!]
    \centering
    \includegraphics[width=0.9\linewidth]{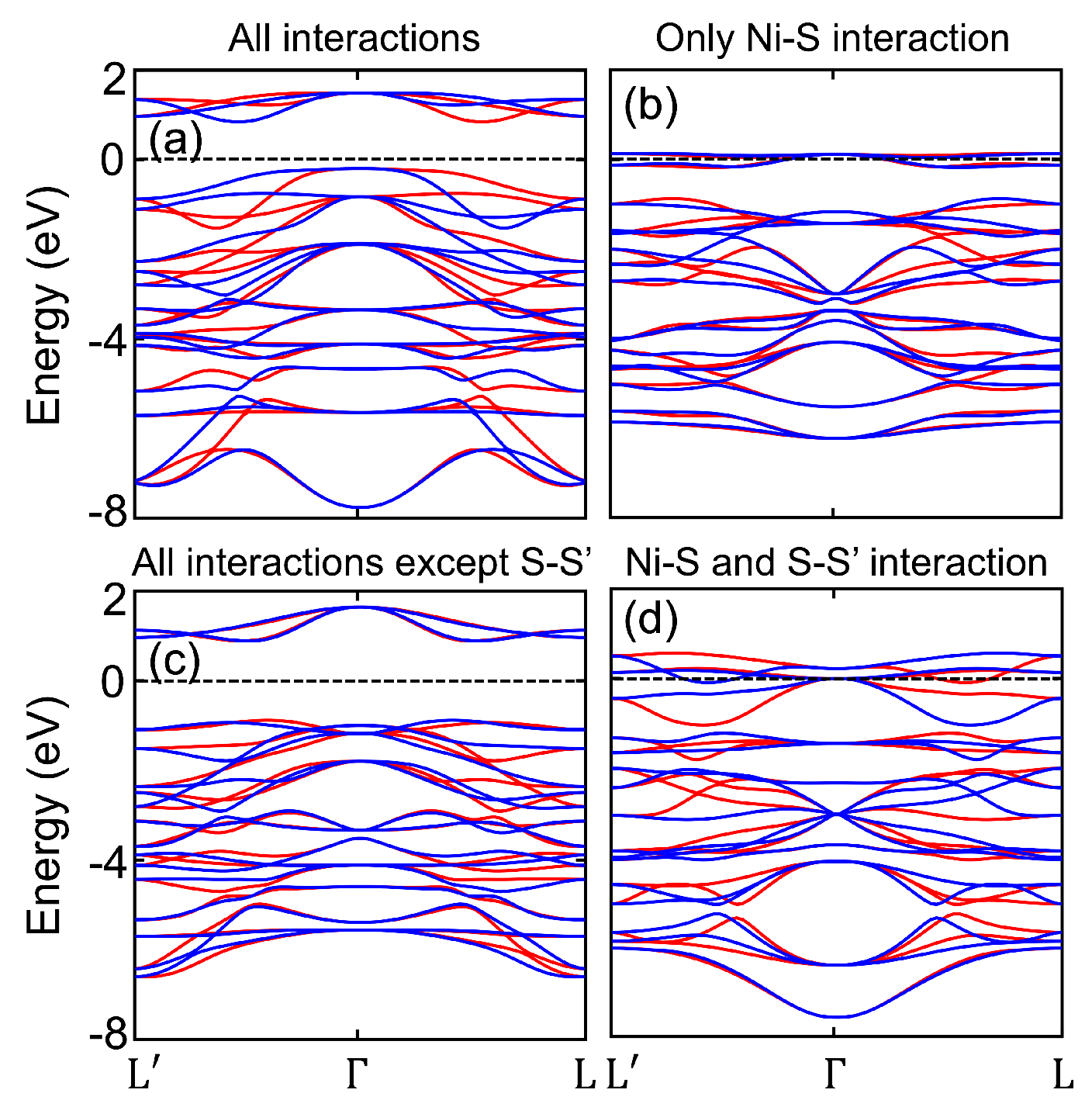}
    \caption{WTB band structure involving all Ni-$d$ and S-$p$ orbitals in the presence of (a) all interactions up to 4th neighbors, (b) nearest neighbor Ni-S interactions only, (c) all interactions except nearest neighbor S-S$^{\prime}$, (d) all nearest neighbor Ni-S and S-S$^{\prime}$ interactions. It implies that in the multi-orbital framework the exclusion of S-S$^{\prime}$ interction reduces the AMSS significantly while the inclusion of the same produces the AMSS matching with the one obtained from DFT.
}
    \label{multi-orb1}
\end{figure}

In Fig. \ref{dft-wann}, we show the band structure obtained from DFT, full Wannierization, and WTB with interactions up to 4th nearest neighbor atoms, as is the case for SKTB (see section \ref{TB-sec}). From the figure, one can see that the DFT bands and Wannier bands are in excellent agreement. Further, we find that the interactions up to the 4th neighbor ($\approx 10 \AA$)  atoms not only captures all the salient features of AMSS, but also matches reasonably well with the DFT bands. Therefore, we consider up to 4th neighbor interactions for further analyses. 

\vspace{-0.5cm}
Fig. \ref{single-orb1} and \ref{single-orb2} depict the band structure for the single orbital model, i.e., one $d$ orbital from Ni atom and one $p$ orbital from S atom (as explained in Fig. 6, 7 and case I, II of section \ref{TB-sec} of the main text) form the bases of the Hamiltonian.
From the figures one can see the Ni-S interactions along with antiferromagnetic Hund's coupling energy cannot create the AMSS. When the S-S$^{\prime}$ interaction is switched on, the altermagnetic band dispersion appears.

For the case of multi-orbital bands, although Ni-S interactions induce altermagnetism due to orbital interplay, the AMSS strength is very weak (see Fig. \ref{multi-orb1}(b)). In fact, all interactions nearest neighbor S-S$^{\prime}$ together are unable to produce the actual AMSS strength (see Fig. \ref{multi-orb1}(c)). Only S-S$^{\prime}$ interaction can produce the real AMSS strength as observed from Fig. \ref{multi-orb1}(a) and (d). Therefore, the results from WTB are consistent with those of SKTB.

%%%%%%%%%%%%%%%%%%%%%%

%\bibliography{ref.bib}
%apsrev4-2.bst 2019-01-14 (MD) hand-edited version of apsrev4-1.bst
%Control: key (0)
%Control: author (8) initials jnrlst
%Control: editor formatted (1) identically to author
%Control: production of article title (0) allowed
%Control: page (0) single
%Control: year (1) truncated
%Control: production of eprint (0) enabled
%

\end{document}